\documentclass[british]{article}

\pdfoutput=1

\usepackage{lmodern}

\usepackage[T1]{fontenc}
\usepackage[utf8]{inputenc}
\usepackage{booktabs}
\usepackage{url}
\usepackage{amsmath}
\usepackage{amssymb}
\usepackage{graphicx}
\usepackage{esint}

\makeatletter

\let\SF@@footnote\footnote
\def\footnote{\ifx\protect\@typeset@protect
    \expandafter\SF@@footnote
  \else
    \expandafter\SF@gobble@opt
  \fi
}
\expandafter\def\csname SF@gobble@opt \endcsname{\@ifnextchar[
  \SF@gobble@twobracket
  \@gobble
}
\edef\SF@gobble@opt{\noexpand\protect
  \expandafter\noexpand\csname SF@gobble@opt \endcsname}
\def\SF@gobble@twobracket[#1]#2{}
\providecommand{\tabularnewline}{\\}

\usepackage{jcappub}
\numberwithin{equation}{section}

\usepackage{fullpage}
\usepackage{array}
\allowdisplaybreaks

\@ifundefined{showcaptionsetup}{}{%
 \PassOptionsToPackage{caption=false}{subfig}}
\usepackage{subfig}
\makeatother

\usepackage{babel}

\begin{document}

\title{\texttt{hi\_class}: Horndeski in the Cosmic Linear Anisotropy Solving
System}

\subheader{\includegraphics[width=0.3\textwidth]{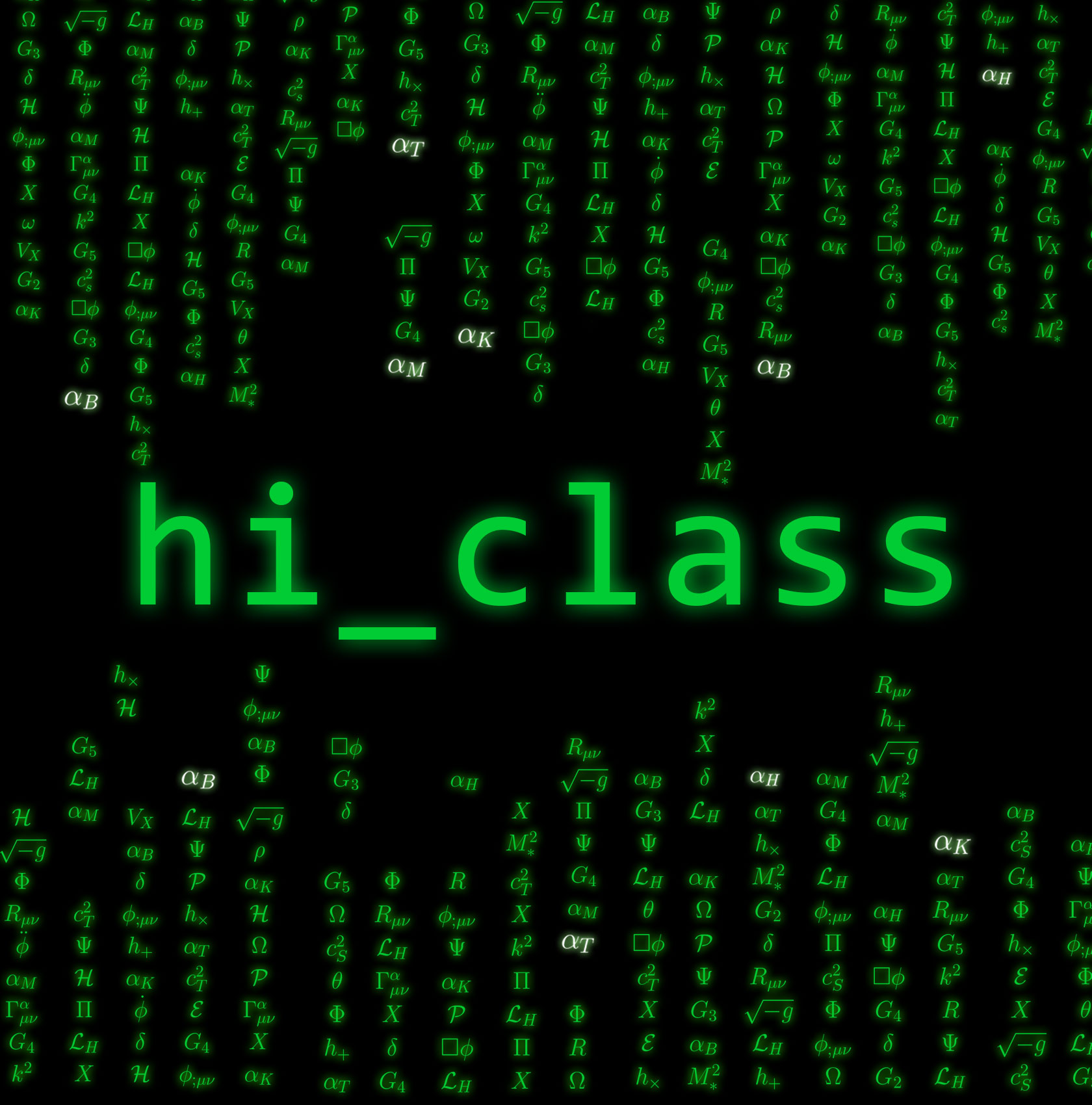} }

\author[a,b,c]{Miguel Zumalacárregui,}

\author[d,e]{Emilio Bellini,}

\author[f]{Ignacy Sawicki,}

\author[g]{\\ Julien Lesgourgues}

\author[e]{ and Pedro G. Ferreira}

\affiliation[a]{Nordita, KHT Royal Institute of Technology and Stockholm University\\
Roslagstullsbacken 23, SE-106 91 Stockholm, Sweden}

\affiliation[b]{Berkeley Center for Cosmological Physics, LBL and University of California at Berkeley\\
Berkeley, California CA94720, USA}

\affiliation[c]{Institut für Theoretische Physik, Ruprecht-Karls-Universität Heidelberg\\
Philosophenweg 16, 69120 Heidelberg, Germany}

\affiliation[d]{Institut de Ciènces del Cosmos, Universitat de Barcelona, IEEC-UB\\
Martì i Franquè 1, E-08028 Barcelona, Spain}

\affiliation[e]{Astrophysics, University of Oxford, \\ Denys Wilkinson Building, Keble Road, Oxford OX1 3RH, United Kingdom}

\affiliation[f]{Central European Institute for Cosmology and Fundamental Physics,\\Fyzikální ustáv Akademie věd ČR, \\Na Slovance 2, 182 21 Praha 8, Czech Republic}

\affiliation[g]{Institut für Theoretische Teilchenphysik und Kosmologie, RWTH Aachen
University\\
D-52056 Aachen, Germany}

\emailAdd{miguelzuma@berkeley.edu}
\emailAdd{emilio.bellini@physics.ox.ac.uk}
\emailAdd{ignacy.sawicki@fzu.cz}
\emailAdd{lesgourg@physik.rwth-aachen.de}
\emailAdd{pedro.ferreira@physics.ox.ac.uk}

\abstract{We present the public version of \texttt{hi\_class}  (\url{www.hiclass-code.net}), an extension of the Boltzmann code CLASS to a broad ensemble of modifications to general relativity. In particular, \texttt{hi\_class} can calculate predictions for models based on Horndeski's theory, which is the most general scalar-tensor theory described by second-order equations of motion and encompasses any perfect-fluid dark energy, quintessence, Brans-Dicke, $f(R)$ and covariant Galileon models.  \texttt{hi\_class} has been thoroughly tested and can be readily used to understand the impact of alternative theories of gravity on linear structure formation as well as for cosmological parameter extraction.}

\keywords{Horndeski, public Boltzmann code, EFT of DE, modified gravity, dark
energy, cosmology, scalar-tensor}

\maketitle

\section{Introduction}

Numerical codes solving the Boltzmann equation in the presence
of gravity have played a hugely important role in the development of modern cosmology. They are the theoretical and computational backbone for model building and precision constraints of cosmological parameters.  The primary focus has been on standard cosmology -- general relativity with a cosmological constant -- where the modern era of fast, publicly available codes, began with  CMBFAST \cite{Seljak:1996is} and was subsequently improved and extended with the Code for Anisotropies in the Microwave Background
(CAMB) \cite{Lewis:1999bs} and the Cosmic Linear Anisotropy Solving
System (CLASS) \cite{Blas:2011rf}. These codes allow us to describe with
percent-level precision the evolution of 
superhorizon initial conditions, through to the large-scale formation
of structure and the propagation of photons.  The output of these codes -- a set of spatial and angular power spectra as a function of time --  can be used to either constrain cosmological
parameters directly \cite{Ade:2014zfo} or to generate the initial conditions for
simulations of more detailed and smaller-scale physics \cite{Valkenburg:2015dsa}.

In this article, we present \texttt{hi\_class}: \emph{Horndeski
in CLASS, }a publicly available extension of the CLASS Boltzmann-solver
code which allows the user to consistently model the presence of an
additional degree of freedom in the gravitational/dark-energy sector
throughout the history of the universe, properly accounting for its effect
on gravitational fields and matter and consistently connecting the
early universe to the observables in these models. This code can then
be used in the standard Monte Carlo Markov Chain module MontePython \cite{Audren:2012wb}
to obtain constraints on parameters of the modifications of cosmology
as well as to ascertain the degradation of the constraints on standard
precision parameters resulting from our ignorance of the underlying
theory for the dark sector. 

The\texttt{ hi\_class} code has already been used to estimate non-linear
effects such as the shift of the baryon acoustic scale \cite{Bellini:2015oua},
get constraints on scalar-tensor theories, using recent data
\cite{Bellini:2015xja}, investigate relativistic effects on ultra-large
scales and cross-correlations between the cosmic microwave background
(CMB) and large-scale structure (LSS) \cite{Renk:2016olm}.

In this first release of the code, we implement
the necessary changes to allow for the modelling of the dynamics of
all models of dark energy which belong to the class of Horndeski theories.
These theories are the most general models of gravity extended by
a single scalar degree of freedom featuring no more than second-order
equations of motion and universally coupled to matter. This class
is extremely wide and covers the substantial majority of models on
the market, including $f(R)$ and $f(G)$ gravities, quintessence,
perfect-fluid dark energy, Galileons, as well as many more exotic
modifications. \texttt{hi\_class} is under continuous development: 
features both extending the default choice of models and simplifying
for the end-user the introduction of new ones are already in advanced
stages of development and will be made public in due course.

The aim of this article is to present an overview of the
modifications of gravity supported in \texttt{hi\_class} and the way
this support is implemented. Updated and more detailed information
can be found on the website \url{www.hiclass-code.net} and the GitHub
repository \url{https://github.com/miguelzuma/hi_class_public}. We
start with a short review of Horndeski theories and the formulation
we use to describe perturbations in Section \ref{sec:Horndeski}.
We then describe the modifications that have been introduced in CLASS
and explain how to use all the features of the code  in Section \ref{sec:CodeDesk}. The
use of the public version of \texttt{hi\_class} is free to the scientific
community but conditional on the inclusion of references to at least
this article and the CLASS paper \cite{Blas:2011rf}.

\section{Horndeski's theory\label{sec:Horndeski}}

\texttt{hi\_class} is a modification of the CLASS code which fully
calculates the evolution of linear cosmological perturbations in dark
energy/modified gravity (DE/MG) models belonging to the Horndeski
class of theories. In these theories the gravitational
sector contains one tensor and one scalar degree of freedom;
their action was first written down in ref.~\cite{Horndeski}, and
then rediscovered as a generalisation of galileons \cite{Nicolis:2008in}
in ref.~\cite{Deffayet:2011gz,Kobayashi:2011nu}. 

Horndeski theories are specified through the action\footnote{Note that we have changed the normalisation of the functions $G_{i}$
with respect to the usual convention by factoring out Newton's constant
as measured on Earth today, $G_{\text{N}}$, and therefore reducing
the mass dimension of the $G_{i}$ by two. Since the Planck mass can
vary in Horndeski models, see the discussion in section \ref{sub:Background}
for how we resolve the ambiguities. In our convention GR is $G_{4}=1/2$
and the rest zero.}
\begin{equation}
S[g_{\mu\nu},\phi]=\int\mathrm{d}^{4}x\,\sqrt{-g}\left[\sum_{i=2}^{5}\frac{1}{8\pi G_{\text{N}}}{\cal L}_{i}[g_{\mu\nu},\phi]\,+\mathcal{L}_{\text{m}}[g_{\mu\nu},\psi_{M}]\right]\,,\label{eq:action}
\end{equation}
\begin{eqnarray}
{\cal L}_{2} & = & G_{2}(\phi,\,X)\,,\label{eq:L2}\\
{\cal L}_{3} & = & -G_{3}(\phi,\,X)\Box\phi\,,\label{eq:L3}\\
{\cal L}_{4} & = & G_{4}(\phi,\,X)R+G_{4X}(\phi,\,X)\left[\left(\Box\phi\right)^{2}-\phi_{;\mu\nu}\phi^{;\mu\nu}\right]\,,\label{eq:L4}\\
{\cal L}_{5} & = & G_{5}(\phi,\,X)G_{\mu\nu}\phi^{;\mu\nu}-\frac{1}{6}G_{5X}(\phi,\,X)\left[\left(\Box\phi\right)^{3}+2{\phi_{;\mu}}^{\nu}{\phi_{;\nu}}^{\alpha}{\phi_{;\alpha}}^{\mu}-3\phi_{;\mu\nu}\phi^{;\mu\nu}\Box\phi\right]\,,\label{eq:L5}
\end{eqnarray}
where the four Lagrangians $\mathcal{L}_{i}$ encode the dynamics
of the Jordan-frame metric $g_{\mu\nu}$ and the scalar field $\phi$.
They contain four arbitrary functions $G_{i}(\phi,X)$ of the
scalar field and its canonical kinetic term, $2X\equiv-\partial_{\mu}\phi\partial^{\mu}\phi$;
we use subscripts $\phi,X$ to denote partial derivatives, e.g.~$G_{iX}=\frac{\partial G_{i}}{\partial X}$.
This scalar field is a new degree of freedom when compared to general relativity
and our implementation chooses appropriate initial conditions and
then fully tracks its dynamical evolution. We restrict our code to
scenarios where the weak equivalence principle holds, i.e.~the matter,
described by the Lagrangian $\mathcal{L}_{\text{m}}$, is universally
coupled to the metric $g_{\mu\nu}$ and does not have direct couplings
with the scalar.\footnote{Modified gravity models may violate the equivalence principle via
non-linear interactions \cite{Hui:2009kc}. However, these effects
beyond the current scope of the \texttt{hi\_class} code, which is
based on linear perturbation theory.}

This general class of actions includes the majority of universally
coupled models of dark energy with one scalar degree of freedom (see the review \cite{Clifton:2011jh}): quintessence
\cite{Wetterich:1987fm,Ratra:1987rm}, Brans-Dicke models \cite{Brans:1961sx},
k-\emph{essence} \cite{ArmendarizPicon:1999rj,ArmendarizPicon:2000ah},
kinetic gravity braiding \cite{Deffayet:2010qz,Kobayashi:2010cm,Pujolas:2011he},
covariant galileons \cite{Nicolis:2008in,Deffayet:2009wt}, disformal
and Dirac-Born-Infeld gravity \cite{deRham:2010eu,Zumalacarregui:2012us,Bettoni:2013diz},
Chameleons \cite{Khoury:2003aq,Khoury:2003rn}, symmetrons \cite{Hinterbichler:2010es,Hinterbichler:2011ca},
Gauss-Bonnet couplings \cite{Ezquiaga:2016nqo} and models screening
the cosmological constant \cite{Charmousis:2011bf,Martin-Moruno:2015bda}.
Archetypal modified-gravity models such as \emph{$f(R)$} \cite{Carroll:2003wy}
and \emph{$f(G)$} \cite{Carroll:2004de} gravity are within our purview. 

We do not, at this time, include scalar-tensor extensions
beyond Horndeski. These models always contain higher derivatives in
the equations of motion, even though they cancel once all the constraints
are solved, leaving second-order equations for the propagating degrees
of freedom \cite{Zumalacarregui:2013pma,Gleyzes:2014dya} (see also
\cite{Langlois:2015cwa,Crisostomi:2016tcp,Deffayet:2015qwa,DAmico:2016ntq}). Horndeski
models can be extended even further to those which contain higher
derivatives in space but not in time \cite{Creminelli:2006xe,Horava:2009if,Blas:2009yd,Gao:2014soa}.
Modifications of gravity with non-scalar degrees of freedom, e.g.~Einstein-Aether
models \cite{Jacobson:2000xp} or ghost-free massive gravity \cite{deRham:2010ik,deRham:2010kj,Hassan:2011zd}
are not included either.

\subsection{Background Evolution\label{sub:Background}}

The expansion rate of the universe $H$ is determined by the energy
density $\rho_{i}$ of all the components of the universe in the standard
manner,
\begin{equation}
H^{2}=\sum_{i}\rho_{i}\,,\label{eq:Friedmann}
\end{equation}
where we are using the CLASS normalisation for the components of the
energy-momentum tensor of $3\rho_{\text{CLASS}}\equiv8\pi G_{\text{N}}\rho_{\text{standard}}$,
and $G_{\text{N}}$ is fixed to be Newton's constant as measured on
Earth today. This sets the internal units for all dimensionful quantities
in the code to Mpc$^{-1}$. In addition notice that \texttt{hi\_class}
and CLASS, and therefore this note, represents by $H$ the physical-time
Hubble parameter, but the time variable is nonetheless conformal time
$\tau$, with its associated $'$ notation for the derivative. Thus
$a^{\prime}=a^{2}H$. 

The evolution of the energy density for the Horndeski scalar field
is determined by the energy density and pressure (rescaled by $8\pi G_{\text{N}}/3)$,
which are obtained from the action ((\ref{eq:action})) through \cite{DeFelice:2011hq}

\begin{align}
\mathcal{\rho_{\text{DE}}}\equiv & -\frac{1}{3}G_{2}+\frac{2}{3}X\left(G_{2X}-G_{3\phi}\right)-\frac{2H^{3}\phi^{\prime}X}{3a}\left(7G_{5X}+4XG_{5XX}\right)\label{eq:rhoDE}\\
 & +H^{2}\left[1-\left(1-\alpha_{\textrm{B}}\right)M_{*}^{2}-4X\left(G_{4X}-G_{5\phi}\right)-4X^{2}\left(2G_{4XX}-G_{5\phi X}\right)\right]\nonumber \\
p_{\text{DE}}\equiv & \frac{1}{3}G_{2}-\frac{2}{3}X\left(G_{3\phi}-2G_{4\phi\phi}\right)+\frac{4H\phi^{\prime}}{3a}\left(G_{4\phi}-2XG_{4\phi X}+XG_{5\phi\phi}\right)-\frac{\left(\phi^{\prime\prime}-aH\phi^{\prime}\right)}{3\phi^{\prime}a}HM_{*}^{2}\alpha_{\textrm{B}}\label{eq:pDE}\\
 & -\frac{4}{3}H^{2}X^{2}G_{5\phi X}-\left(H^{2}+\frac{2H^{\prime}}{3a}\right)\left(1-M_{*}^{2}\right)+\frac{2H^{3}\phi^{\prime}XG_{5X}}{3a}\,,\nonumber 
\end{align}
where $a$ is the scale factor of the Universe which appears as a
result of the choice of definition of $H$. Note the presence of terms
$\propto H^{2},H^{3}$ stemming from the non-minimal coupling to the
curvature in the action. The above expressions have been simplified
by introducing $M_{*}^{2}$, defined below, and $\alpha_{\text{B}}$,
defined in appendix \ref{sub:appendix_alphas}. 

Cosmological backgrounds in Horndeski theories admit a time-evolving
effective Planck mass, defined as the normalisation of the kinetic
terms of the graviton. Horndeski models thus implicitly contain a
free dimensionless function $M_{*}^{2}(\tau)$ --- which we call the
\emph{cosmological strength of gravity ---} which represents the square
of the ratio of the cosmological Planck mass to the Earth-bound one
and which is determined from the action through
\begin{equation}
M_{*}^{2}\equiv2\left(G_{4}-2XG_{4X}+XG_{5\phi}-\frac{H\phi^{\prime}}{a}XG_{5X}\right)\,.\label{eq:Mstar}
\end{equation}
There are two subtleties which impact today's value of $M_{*}^{2}$: 
\begin{itemize}
\item The measurement of $G_{\text{N}}$ using Cavendish-like experiments
in the context of a Horndeski gravity in fact measures the sum of
the gravitational and the fifth forces and thus not quite the local
value of the Planck mass. However, as the Solar-System constraints
on the $\beta$ and $\gamma$ PPN parameters are of the order 1 part
in $10^{5}$ \cite{Will:2014kxa}, the difference between $G_{\text{N}}$
and the underlying $M_{\text{Pl}}^{-2}$ is negligibly small for the
purposes of cosmological constraints and we neglect it.
\item The non-linear screening of gravitational interactions (e.g.~the
chameleon \cite{Khoury:2003aq} or Vainshtein \cite{Vainshtein:1972sx}
mechanisms) changes the strength of gravity between cosmological scales
and the Solar System (it essentially is a change to the background
solution, i.e.~the local value of $\phi$ and $\phi'$ compared to
these values in cosmology today). Thus Cavendish-like experiments
determine $M_{\text{Pl}}$ locally but this is not necessarily its
value at cosmological scales today. Thus \emph{if }there is a screening
mechanism active, the cosmological strength of gravity today, $M_{*,0}^{2}$,
is a free parameter to be constrained by observations. If no screening
is active for the Solar System, $M_{*,0}^{2}=1$ up to the correction
in the first bullet point and the initial conditions for $M_{*}^{2}$
should be set in such a way so as to achieve this. Note though that
in the no-screening case, PPN constraints can be directly applied
to cosmological scales and imply that the properties of gravity are
very similar to GR, presumably preventing any significant modifications
to cosmological observables.
\end{itemize}
We emphasise that the cosmological evolution does not actually depend
on Solar-System measurements, but rather $G_{\text{N}}$ sets the
units necessary to convert the measurement of the dimensionful CMB
temperature to an effect on space-time curvature that the radiation
produces and thus the meaning that we ascribe to the Mpc distance
unit and all dimensionful quantities in cosmological observations.

As a result of the evolving Planck mass, there also exists an ambiguity
in the definition of the energy-momentum tensor. The particular choice
of variables made here in eqs.~(\ref{eq:rhoDE}-\ref{eq:pDE}) allows
us to write down the conservation equation for the DE energy density
and the equation of state in the standard manner
\begin{align}
\rho_{\textrm{DE}}^{\prime} & =-3aH\left(\rho_{\text{DE}}+p_{\text{DE}}\right)\,,\label{eq:EnCons}\\
w_{\textrm{DE}} & \equiv\frac{p_{\text{DE}}}{\rho_{\text{DE}}}\,.\nonumber 
\end{align}
This choice also implies that the density of e.g.~pressureless
matter scales as in the standard case, $\rho_{\text{m}}\propto a^{-3}$,
despite the evolving masses. This matches the implicit assumption in
all published observational constraints on $w_{\text{DE}}$. Note
though that the equation of state $w_{\textrm{DE}}$ would not necessarily
be the ratio of pressure and energy density as measured by a comoving
observer in the past if they had the means to probe the instantaneous
energy density and pressure (see the discussion in Appendix \ref{sec:AppBack}).

The flexibility of choosing arbitrary functions $G_{i}(\phi,X)$ and
initial conditions for the scalar field means that essentially any
choice of the function $w_{\text{DE}}(\tau)$ can be made. It is only
the choice of the history of $\rho_{\text{DE}}(\tau)$ (equivalently
$w_{\textrm{DE}}(\tau)$ and the density today $\rho_{\text{DE,0}}$)
that impacts the evolution of the background and thus specifying these
is enough to describe all possible cosmological backgrounds. The only
theoretical constraint is that such a background does not suffer from
instabilities. We discuss this issue in Section \ref{sec:Stability}. 

Specifying the evolution of the background does
not, however, determine the evolution of perturbations. Implementing the evolution
of perturbations properly was our aim in writing \texttt{hi\_class},
since it is discriminating between the various perturbation
evolutions possible on the same observed background that allows us
to differentiate between the models of modified gravity.

\subsection{Description of Perturbation Dynamics\label{sub:alphas}}

The primary focus of Einstein-Boltzman solvers is linear cosmological perturbations on a homogeneous and isotropic background. The evolution equations that completely describe the dynamics of the perturbed fields can be obtained from the linearized field equations arising from the action in Equation \eqref{eq:action}. An alternative, and enlightening, approach is to expand the action in Equation \eqref{eq:action} to second order in linear perturbations of $g_{\alpha\beta}$, $\phi$ and the remaining matter fields. The action for perturbations consists of a sum of terms, quadratic in the perturbation fields, each of which is multiplied by a time dependent function which solely depends on the background cosmology. One can choose a basis for these time dependent coefficients and a particularly useful one was given in \cite{Bellini:2014fua}, since they relate most directly to the physical
observables and to the propagating degrees of freedom in the Horndeski
models. Once the
background $H(\tau)$ is specified through Eq. (\ref{eq:Friedmann}),
the complete set of possible modifications of Einstein's equations
describing the evolution of linear perturbations in Horndeski theories
is determined by four functions of time: the dimensionless
cosmological strength of gravity $M_{*}^{2}(\tau)$ and three $\alpha_{i}(\tau)$.
In this document we will refer to these four functions collectively
as ``$\alpha$ functions''. The definitions of the $\alpha$ functions
in terms of the action are given in Appendix \ref{sub:appendix_alphas}.

Working with the quadratic action arising from Equation \ref{eq:action} also allows us to connect with  the approach of Effective Field Theory of Dark Energy (EFT) \cite{Gubitosi:2012hu,Bloomfield:2012ff,Gleyzes:2013ooa}.
There the idea is to encode all the possible modifications of Einstein equations
consistent with the symmetries of the cosmological background and
gauge freedom in terms of coefficients which are functions of time
only, and which multiply fixed operators which carry the information
on scale dependence. In the EFT approach, a choice of basis for the operators must
also be made (see ref.~\cite{Bellini:2014fua} for a translation of
the EFT of DE operators into the $\alpha$ function basis). 

\texttt{hi\_class} solves the equations for the tensor (\ref{eq:tesors})
and scalar modes (\ref{eq:metric_00}-\ref{eq:metric_vx}) of the
modified gravitational sector, together with the Boltzmann hierarchy
for the standard matter components, to give the full predictions for
linear large-scale structure. The four functions describing the modification
of gravity can be divided into two pairs: the first related to both
the scalar and the tensor modes, the second only to the scalar. We
refer the reader to ref.~\cite{Bellini:2014fua} for details of the
physical impact of these parameters.
\begin{enumerate}
\item Non-minimal coupling of gravity. These functions modify both the scalar
and the tensor propagation: 

\begin{enumerate}
\item $M_{*}^{2}(\tau)$\emph{, the cosmological strength of gravity. }$M_{*}^{2}$
is the dimensionless product of the normalisation of the kinetic term
for gravitons and $8\pi G_{\text{N}}$ as measured on Earth. It thus
encodes the difference between the solar system and cosmology in the
gravitational force/space-time curvature produced by a fixed amount
of energy . Large-scale structure is sensitive only to the time variation
of the Planck mass,
\begin{equation}
\alpha_{\textrm{M}}\equiv\frac{\mathrm{d}\ln M_{*}^{2}}{\mathrm{d}\ln a}\,,\label{eq:running def}
\end{equation}
or the \textit{Planck-mass run rate}. However, as a result of screening,
it is possible that Newton's constant as measured by local experiments,
$G_{\text{N}}$, and that on cosmological scales are different. Thus
only if screening in the Solar System is active, the value today of
$M_{*,0}^{2}$ is free and largely unconstrained. If the Solar-System
is unscreened, $M_{*,0}^{2}=1$. Also note that Planck observations
of the recombination history imply that the value of the Planck mass
at recombination cannot differ from that measured in the Solar System
by more than 1\% \cite{Ade:2014zfo}. 
We give users the choice to employ as the principal parameterisation:
either $M_{*}^{2}(\tau)$, or $\alpha_{\text{M}}(\tau)$ together
with the initial value of $M_{*}^{2}$.

\item $\alpha_{\textrm{T}}(\tau)$, \textit{tensor speed excess}. This parameter
denotes the difference in the propagation speed of gravitational waves
compared to the speed of light, i.e.~$\alpha_{\textrm{T}}=c_{\textrm{T}}^{2}-1$.
It applies to modes propagating on cosmological scales and is currently
the most weakly constrained parameter from physics other than cosmology
(see ref.~\cite{Blas:2016qmn} and references therein). 

In Horndeski theories one expects screening of high-density environments
like the Earth or the location of a system that could produce gravitational
waves (GWs) and thus no modification in the production or detection
of GWs is likely \cite{deRham:2012fw,deRham:2012fg} (although see
ref.~\cite{Jimenez:2015bwa}). The GW dispersion relation is only
modified by the change in the sound speed and thus one would not expect
any frequency-dependent modifications of the propagation speed. An
observation of an indisputable electromagnetic counterpart to a gravitational
wave event at cosmological distances would put a constraint on this
parameter tight enough to render the remaining uncertainty irrelevant
for structure formation \cite{Lombriser:2015sxa,Bettoni:2016mij}.
\end{enumerate}

Note that either $\alpha_{\text{M}}$ or $\alpha_{\text{T}}$ must
not vanish in order for gravitational slip to be generated \cite{Saltas:2014dha}
(e.g.~modifying the spatial-traceless metric equation \ref{eq:metric_ij}).
In such a case, the equation of motion for the propagation of gravitational
waves (\ref{eq:tesors}) is also modified.

\item Kinetic terms. The scalar mode is also affected by the following
two functions:

\begin{enumerate}
\item $\alpha_{\textrm{B}}(\tau)$, \textit{braiding}. This operator gives
rise to a new mixing of the scalar field and metric kinetic terms
for the propagating degree of freedom \cite{Bettoni:2015wta}. This leads to a modification
of the coupling of matter to the curvature, independent and additional
to any change in the Planck mass. This is typically interpreted as
an additional fifth force between massive particles and can be approximated
as a modification of the effective Newton's constant for perturbations.
It is present in archetypal modified gravity models such as Brans-Dicke and $f(R)$
gravity (see \cite{Bellini:2014fua} for details). A purely conformal
coupling of the scalar to gravity leads to the universal property
$\alpha_{\text{M}}+\alpha_{\text{B}}=0$. 
\item $\alpha_{\textrm{K}}(\tau)$, \textit{kineticity}. Coefficient of
the kinetic term for the scalar d.o.f.~before demixing (see ref.~\cite{Bellini:2014fua}).
Increasing this function leads to a relative increase of the kinetic
terms compared to the gradient terms and thus a lower sound speed
for the scalar field. This creates a sound horizon smaller than the
cosmological horizon: super-sound-horizon the scalar does not have
pressure support and clusters similarly to dust. Inside, it is arrested
and eventually can enter a quasi-static configuration \cite{Sawicki:2015zya}.
When looking only at the quasi-static scales, inside the sound horizon,
this function cannot be constrained \cite{Gleyzes:2015rua}.%
\footnote{Constraints on $\alpha_{\rm K}$ can be obtained from observations on ultra-large scales \cite{Alonso:2016suf}.}
This is the only term present in the simplest DE models, e.g.~quintessence
and in perfect-fluid dark energy.
\end{enumerate}

Note that the quantity $D\equiv\alpha_{\textrm{K}}+\frac{3}{2}\alpha_{\textrm{B}}^{2}$
has to be \emph{strictly }positive at all times to avoid strong coupling
problems or ghosts. This comes from the fact that $D$ represents
the demixed kinetic term of the additional scalar degree of freedom.
Moreover $D=0$ is a pressure singularity and cannot be crossed in
background dynamics, even though this might seem possible at the level
of parameterised perturbations (see e.g.~ref.~\cite{Easson:2011zy}
for example phase-space trajectories in cosmology).

\end{enumerate}
The Horndeski class of theories together with the freedom to choose
initial conditions is large enough to guarantee that there exists
some action that would result in any choice of equation of energy
density of dark energy $\rho_{\text{DE}}(\tau)$ and an arbitrary
choice of the $\alpha$ functions, apart from the exception mentioned
above. Such an arbitrary choice may be unstable and/or fine tuned and might thus
not be a good solution in practice. Conversely, the only information
relevant for linear Boltzmann codes that a particular model based
on a full covariant action provides is exactly contained in this evolution
history of the energy density of the dark energy and the $\alpha$
functions for that model. It is enough to calculate these as functions
of time to then be able to obtain the predictions of that model at
all scales with no loss of information. We can thus take two approaches
to modelling: 
\begin{enumerate}
\item Start by choosing a full covariant classical action within the Horndeski
class according to some overarching principle. Then calculate the
self-consistent predictions for this particular model for the background
and the perturbations, given some initial conditions for the background
scalar. This allows the user to constrain the values of particular
parameters of the model. 
\item Use a specification of some evolution history for the cosmological
background and the $\alpha$ functions as the model, without further
reference to the action. This allows the user to test whether there is any evidence for the departure
of the behaviour of growth from the concordance predictions and what
sort of physical properties of the gravity theory could be driving
that growth.
\end{enumerate}
Our implementation in \texttt{hi\_class} allows us to cover both the
approaches with the same code. Approach (1) requires an additional
pre-computation of the background and the $\alpha$-functions as part
of an initialisation module which then feeds these quantities to the
same code as (2) would use (see Figure \ref{fig:code-structure}).
For the initial roll-out, we are enabling the parameterised EFT approach
(2) in this version of \texttt{hi\_class}, but will be bringing
an automatised implementation beginning with a full covariant
action in the full release.

\subsection{Stability Conditions\label{sec:Stability}}

When specifying a model, it is possible to choose a set of $\alpha$ functions which lead
to exponentially unstable perturbations. This is all the more likely when choosing a set of $\alpha$s which are completely unrelated to a legitimate scalar field model, but can also happen for what seems like a valid background but which might actually have a kinetic
term with the wrong sign (e.g.~forcing a perfect fluid to evolve with $w_{\textrm{DE}}<-1$).
Every background supplied to \texttt{hi\_class }is tested to make
sure that no such short-timescale instabilities are present and the
background can be trusted over cosmological timescales. Only if such
stability tests are passed, are the perturbation equations solved
for that set of parameters.

There are essentially three types of instabilities that would disqualify
a choice of background. 
\begin{itemize}
\item\emph{Ghost instabilities} occur when the
sign of the kinetic term of a degree of freedom is wrong, giving negative
energy modes and violating unitary evolution. If the mass of these
ghosts is less than the cutoff of the theory, this would lead to rapid
pair production into negative-energy modes, destroying the background.
\item
\emph{Gradient instabilities} occur when the sound speed squared is
negative. This leads to an exponentially growing instabilities with
a rate corresponding to the shortest mode allowed by the effective
theory. Importantly, such instabilities may not be visible until very
short modes and/or higher-order (interaction) terms are included in
the evolution and thus a seemingly unsuspect evolution of linear perturbations
is not sufficient that the predictions are at all correct, necessitating
the stability tests.
\item
\emph{Tachyon instabilities} occur when the mass squared for the perturbations is negative. This leads to a power-law instability at \emph{large} scales, which comes under control when the mode enters the sound horizon. Since this has a significant on the calculated observables, we choose to let the data exclude such cases, rather than creating a hard stability test.
\end{itemize}

The Horndeski class of models affects the evolution of both scalars
and tensors. \texttt{hi\_class} thus performs four tests to verify
the stability of the background:

\begin{align}
Q_{\textrm{S}}= & \frac{2M_{*}^{2}D}{\left(2-\alpha_{\textrm{B}}\right)^{2}}>0\,,\qquad\qquad D\equiv\alpha_{\textrm{K}}+\frac{3}{2}\alpha_{\textrm{B}}^{2}>0\label{eq:scalar_ghost}\\
c_{\text{s}}^{2}= & \frac{1}{D}\left[\left(2-\alpha_{\textrm{B}}\right)\left(-\frac{H^{\prime}}{aH^{2}}+\frac{1}{2}\alpha_{\textrm{B}}\left(1+\alpha_{\textrm{T}}\right)+\alpha_{\textrm{M}}-\alpha_{\textrm{T}}\right)-\frac{3\left(\rho_{\textrm{m}}+p_{\textrm{m}}\right)}{H^{2}M_{*}^{2}}+\frac{\alpha_{\textrm{B}}^{\prime}}{aH}\right]>0\label{eq:scalar_grad}\\
Q_{\textrm{T}}= & \frac{M_{*}^{2}}{8}>0\label{eq:tensor_ghost}\\
c_{\text{T}}^{2}= & 1+\alpha_{\textrm{T}}>0\,,\label{eq:tensor_grad}
\end{align}
$Q_{\textrm{S}}$ and $Q_{\textrm{T}}$ respectively represent the
kinetic terms of scalar and tensor sectors after demixing; $c_{\text{s}}^{2}$
and $c_{\text{T}}^{2}$ are the sound speeds of these two degrees
of freedom. 

We should note that it is still possible that a background that
passes the above tests, is nonetheless not healthy as a result of
some non-linear instability, for example when the null energy condition
is violated \cite{Sawicki:2012re}. On the other hand, one may also
argue that the appearance non-linear structure from the exponential
instabilities at small scales might arrest the destabilisation and
save a choice of a background that does not satisfy the stability
checks. Strictly speaking, such a background would no longer be FRW
and therefore the description in terms of the $\alpha$ functions
would no longer be complete, but it is possible that the largest scales
remain close enough to FRW to allow a description that we have implemented.\footnote{But this approximate background might have different effective values
of the $\alpha_{i}$ to the original ones we started with, presumably
ones implying stability.} In any case, should such a scenario be favoured by the user, or the
user would like to check which part of the parameter space is constrained
on theoretical grounds as opposed to just data on large scales, we
give them the freedom to relax or disable these checks entirely. See
\cite{Salvatelli:2016mgy,Perenon:2016blf,Peirone:2017lgi} for studies
of the impact of stability conditions on parameterized models.

\subsection{Initial Conditions\label{sub:ICs}}

The formulation of perturbation equations that we have implemented
depends on using as the variable for the fluctuation of the scalar
field a velocity potential, 
\begin{equation}
V_{X}\equiv a\frac{\delta\phi}{\phi^{\prime}}\,.\label{eq:VXdef}
\end{equation}
Its dynamics is described by the second-order differential equation (\ref{eq:metric_vx}),
which requires two initial conditions, to be solved. 

The chosen initial conditions are specified superhorizon and typically evolve rapidly into the natural attractor of the theory, at least in the case of negligible
modified gravity effects at early times (for the discussion of perfect
fluid DE see \cite{Ballesteros:2010ks}). Indeed, we have found that, for parameterisations enabled in this version of \texttt{hi\_class}, the choice of initial conditions is not material to the study of the late universe. We have thus included a couple of simple choices to allow the user to verify the possible impact of initial conditions. Moreover, the code includes a flexible structure to allow the user to easily incorporate other choices.

The simplest possible choice is to set both the field perturbation and its derivative to zero

\begin{equation}
V_{X}(\tau_{0})=0\,,\quad V_{X}^{\prime}(\tau_{0})=0\,.\label{eq:ic_zero}
\end{equation}

In simple inflationary models the initial perturbations are given
by \emph{single-clock} initial conditions. In the case of the scalar
field this scenario is implemented by assuming $\delta\phi(\tau,\vec{x})=\phi\left(\tau+\delta\tau(\tau,\vec{x})\right)$,
i.e.~it is given by the background value of the scalar field evaluated
on perturbed constant time hypersurfaces. In terms of the field perturbation
and relating $\delta\tau$ to the perturbation in the photon density
yields
\begin{equation}
V_{X}(\tau_{0})=-\frac{1}{4H}\frac{\delta\rho_{\gamma}}{\rho_{\gamma}}\,,\quad V_{X}^{\prime}(\tau_{0})=0\,.\label{eq:ic_single_clock}
\end{equation}

The detailed initial conditions for Horndeski theories will be presented with the full version of the code.

\section{Description of the Code\label{sec:CodeDesk}}

\texttt{hi\_class }is a forked version of the Cosmic Linear Anisotropy
Solving System (CLASS), a modern Boltzmann-Einstein code designed
to compute cosmological predictions in the linear regime. CLASS has
been developed from the outset to make it intuitive and simple to
modify. It consists of several modules that perform tasks sequentially
(reading parameters, computing the background, thermodynamics, perturbations,
etc.; see \cite{Lesgourgues:2011re}), with each module relying on
the results generated by the previous ones. A major design goal of
the code has been flexibility, so CLASS can easily incorporate non-standard
cosmological models. Several extensions of the $\Lambda$CDM concordance
scenario, such as massive neutrinos, warm dark mater, curvature \cite{Lesgourgues:2011rh,Lesgourgues:2013bra,Tram:2013ima},
but also quintessence and perfect-fluid dark energy, have already
been implemented in a manner compatible with the CLASS principles.
These, as well as the standard components (photons, baryons, dark
matter) are treated sequentially so that a block of instructions (to
account for each component) is executed only if such a component is
included in the model and ignored otherwise. This allows CLASS to
remain both fast, flexible and self-consistent in the presence of
many different modifications. 

\texttt{hi\_class} incorporates general Horndeski scalar-tensor modifications
of gravity as yet another option in this scheme, allowing the user
to investigate with ease the theory of gravity governing the Universe.
Moreover, \texttt{hi\_class} allows the user to investigate the new
degeneracies that appear when modifications of gravity are consistently
included together with other extensions of the concordance model.
\texttt{hi\_class} can be used with other cosmic components such as
massive neutrinos, but not yet spatial curvature. Other features of
CLASS such as the option to compute relativistic corrections to observables
(e.g.~the CLASSgal code \cite{DiDio:2013bqa}) are also compatible
\cite{Renk:2016olm}.

\begin{figure}
\begin{centering}
\includegraphics[width=0.9\textwidth]{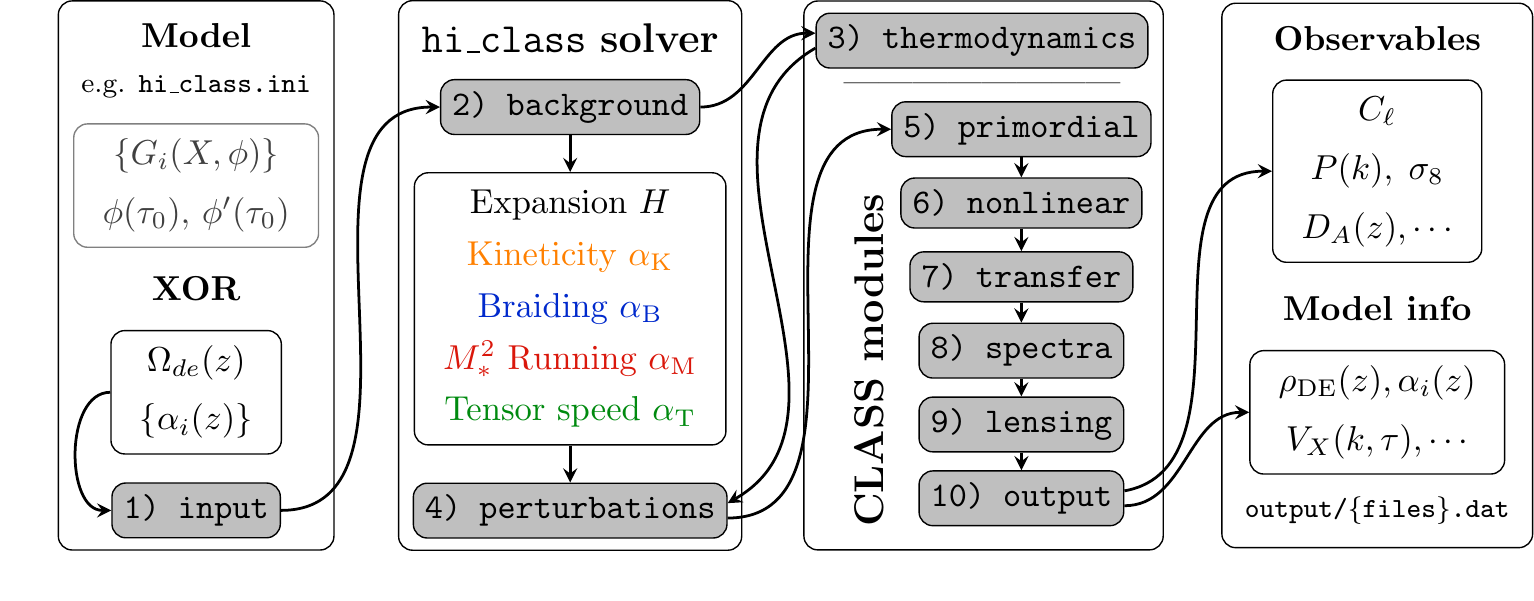}
\par\end{centering}

\caption{\texttt{hi\_class} structure. The arrows represent the flow of information
between modules, while the numbers represent the order of execution
of the modules. The user has to specify a parameterisation for the
expansion and the $\alpha$ functions (in the full version, a model
based on a full covariant action is also possible) that is interpreted
by the input module. Then the background is computed giving the Hubble
rate and the $\alpha$ functions. The unmodified thermodynamics module
computes the thermal evolution using this background expansion history.
The code then uses these as inputs to determine the evolution of the
perturbations. The rest of the code is unmodified with respect to
the public version of CLASS. \texttt{hi\_class }can be used to output
both cosmological observables as well as details about the model and
its evolution. \label{fig:code-structure}}
\end{figure}

By using the description of the dynamics in terms of the $\alpha$
functions, as laid out in section \ref{sec:Horndeski}, we have been
able to write a common routine for all Horndeski models, leaving the
user with the freedom to determine the particular choice of model,
which boils down to a description of the evolution history of the
cosmological background and the $\alpha$ functions, but not to concern
themselves with the numerical algorithm. From the perspective of \texttt{hi\_class},
it is immaterial if this evolution history originates from a model
based on a full covariant action or if it is generated by assuming
a parameterisation. In both cases, the code pre-computes the appropriate
quantities from the given model automatically and evolves the predictions
for large-scale structure (see Figure \ref{fig:code-structure}).

We describe the general algorithm used by \texttt{hi\_class} in section
\ref{sub:Implementation}, discuss the choice of models we are releasing
to the public at this juncture in section \ref{sub:Models} and demonstrate
the impact on observables of some chosen modifications of gravity
in the Horndeski class in section \ref{sub:Plots}.

\subsection{Implementation of Horndeski Dynamics\label{sub:Implementation}}

The code first runs the background module, solving for the background
evolution. The scale factor as a function of conformal time is obtained
by integrating the Hubble rate as given by the Friedmann equation
(\ref{eq:Friedmann}), starting deep in the radiation era. The Hubble
parameter depends on the matter components (DM, baryons, radiation,
massive neutrinos) as well as the model for the energy density of
the Horndeski scalar $\rho_{\text{DE}}(\tau)$ and its corresponding
pressure $p_{\text{DE}}(\tau)$ specified by the user.

The user can specify either a parameterization for the density and
pressure, possibly with multiple parameters (e.g.~$\Omega_{\text{DE},0}$,
$w_{0},$$w_{a}$) or by specifying a model based on an action (not
currently available in the public version). In the latter case, the
energy density and pressure typically depend indirectly on the parameters
of the action through the evolution of an underlying scalar field
value and also its initial conditions at the background level. \texttt{hi\_class}
employs CLASS's automatic routine which iteratively solves for the
evolution, making sure that the dark energy reaches the appropriate
value for its density fraction today as required by the closure relation
\begin{equation}
\sum_{i}\Omega_{i,0}=1\,.\label{eq:Closure}
\end{equation}
The user is required to specify which of the background parameters
is to be varied by the code to give the correct density fraction.
Note that the degrees of freedom of the background parameterisation
are always one fewer than the number of parameters, since the constraint
(\ref{eq:Closure}) must always be satisfied: e.g.~in flat $\Lambda$CDM,
once the density fraction of CDM, baryons and radiation are specified,
$\Omega_{\Lambda}$ is a derived parameter and there is no remaining
freedom.

At this point, all background quantities — those that depend only
on time (including indirectly through other background functions),
but do not depend on scale — are computed and stored in an interpolation
table for further use in subsequent modules. This includes the energy
density and pressure of all matter species, including the Horndeski
scalar field, as well as the $\alpha$ functions (as prescribed by
the model chosen by the user) and all relevant combinations of quantities
that enter the coefficients of the perturbation equations (see section
\ref{sub:Synchronous_perturbations}). Whenever these quantities depend on time derivatives (e.g.~of the
$\alpha$ functions), they are computed numerically at this stage.
For instance eq. (\ref{eq:Mstar}) defines the running of the effective
Planck mass. 

For parameterisations in which it is the cosmological
strength of gravity $M_{*}^{2}(\tau)$ that is supplied by the user,
$\alpha_{\text{M}}$ is obtained through a numerical derivative. On
the other hand, whenever $\alpha_{\text{M}}$ is supplied by the user,
$M_{*}^{2}$ is obtained by the background module using numerical
integration, requiring an input of an initial value of $M_{*}^{2}$;
this procedure gives the appropriate value of the cosmological strength
of gravity today $M_{*,0}^{2}$ as a derived quantity. We note, however,
that with the choice of definition for $\rho_{\text{DE}}$ and $p_{\text{DE }}$,
eqs~(\ref{eq:rhoDE}-\ref{eq:pDE}), $M_{*}^{2}$ has no effect on
background evolution since it can always be reabsorbed into an appropriate
rescaling of both the Hubble constant $H_{0}$ and the physical energy
densities of all the components of the Universe. It is only when perturbations
are considered that it plays a role.

The code then moves on to evaluate the stability conditions given
in section (\ref{sec:Stability}). This allows us to check if the
particular background requested by the user is stable and consistent
for the particular model of gravity chosen. If any of the conditions
fails to be satisfied, the code interrupts execution and returns an
error. If running on an MCMC, the sampler assigns a negligible likelihood
to the parameters, indicating that the model is incompatible with
data, and moves on to try different values. The user has the freedom
to override this consistency check by setting alternative thresholds
for each of the stability conditions (\ref{eq:scalar_ghost}-\ref{eq:tensor_grad})
or skip this test altogether. Note that $M_{*,0}^{2}$ enters the
stability conditions and is as of this point a physically meaningful
quantity.

The perturbations module runs over the chosen range of scales $k$,
solving the modified equations for the metric (\ref{eq:metric}) and
scalar-field in the synchronous gauge (c.f. appendix \ref{sub:Synchronous_perturbations}).
Numerical integrations are performed for the metric potential $\eta$
through the first-order equation (\ref{eq:metric_0i}) and the scalar
field eq. (\ref{eq:metric_vx}), while other quantities are determined
algebraically from the remaining gravitational eqs (\ref{eq:metric_00},\ref{eq:metric_ii},\ref{eq:metric_ij}).
The effect of universally coupled Horndeski models is to modify just
the gravitational sector. The effect on the dynamics of the matter
components is solely through the modification of the gravitational
potentials and thus the matter perturbations are evolved using the
perturbed continuity/Euler equations or the Boltzmann hierarchy as
appropriate, with no changes to the methods implemented in CLASS. 

The initial conditions for all components are set in an early epoch
that is both in the radiation era and guarantees that the scale in
question is super-horizon, i.e.~$k\tau\ll1$ (see ref.~\cite{Blas:2011rf}
for details). \texttt{hi\_class} by default allows for two possible
initial conditions, as described in section \ref{sub:ICs}. As mentioned above, we have
checked that the choice of initial conditions does not affect the
predictions whenever the $\alpha$ functions are negligible at early
times.

For full consistency, we also modify the evolution of the tensor modes
according to eq. (\ref{eq:tesors}). This does not significantly affect
observables unless there is a significant primordial amplitude of
tensor modes \cite{Amendola:2014wma}.

\texttt{hi\_class} retains the full dynamics across cosmic history
and does not invoke any further simplifications to solve the linear-perturbation
equations. However, for some models, the computation of the evolution
of the scalar field value can slow down considerably when the $\alpha$
functions are small and a hierarchy exists between them (in particular
when $\alpha_{\text{K}}\ll\alpha_{\text{B}}^{2}$). This can happen
for some models, typically at early times, and can make computations
unfeasible as part of an MCMC chain. Such an issue is typically caused
by a very large effective mass term for the mode in the equation of
motion for the scalar-field perturbation, leading to very rapid oscillations
in the solution, which the integrator attempts to track. Writing the
equation of motion for the scalar (\ref{eq:metric_vx}) schematically,
\begin{equation}
V_{X}''+\nu aHV_{X}'+\left(a^{2}\mu^{2}+c_{\text{s}}^{2}k^{2}\right)V_{X}= S\,,
\end{equation}
the physical origin of such a problem can be two-fold: either the
speed of sound (\ref{eq:scalar_grad}) is very high (extremely superluminal),
$c_{\text{s }}^{2}\gg1$ , so that modes which are outside of the
cosmological horizon at early times nonetheless are inside the sound
horizon and oscillating. This is probably an undesired feature for
the model. Or: the mass of the scalar field is very high initially,
$\mu\gg H$. In such a case, scales with $a\mu>c_{\text{s}}k$ are
outside of the Compton length of the scalar and the scalar perturbation
$V_{X}$ is negligible. Such a scenario is typical, for example, in
$f(R)$ models which always have $\alpha_{\text{K}}=0$. 

To accelerate the computation in these cases, we have introduced a
user-defined parameter which regularises the kinetic term
of the scalar field by adding a small constant value to the kineticity,
$\alpha_{\text{K}}^{\text{tot}}(\tau)=\alpha_{\text{K}}(\tau)+\alpha_{\text{K}}^{\text{reg}}$.
This artificially reduces the sound speed and mass at early times,
leading the scalar-field to behave similarly to a dust component.
Depending on the model, this might alter the configuration of the
scalar during recombination sufficiently to have an effect on the
predictions. However, we have tested that changing the value of $\alpha_{\text{K}}^{\text{reg}}$
does not significantly affect the results. We leave it to the user
to test what maximal value for $\alpha_{\text{K}}^{\text{reg}}$ can
be used for their chosen model and parameter range of interest without
biasing the predictions. The next version of \texttt{hi\_class} will
incorporate suitable approximation schemes to avoid the necessity
of kineticity regularization.

The output of the perturbations module consists of source
functions dependent on time $\tau$ and scale $k$ that store information
(i.e.~solutions of the perturbed variables and combinations thereof).
This data is stored in memory and then used in subsequent modules
to compute the observables. The CLASS code has been designed to avoid
duplicated equations or implicit assumptions in latter parts of the
code. This implies that the Friedmann equation appears only at one
point in the background module, and similarly for the (modified) Einstein
equations in the perturbations module. These are the only places where
the dynamics of modified gravity enter, and there are essentially
no changes in any subsequent module. Note however that certain options
in CLASS might still be inconsistent with the modified-gravity dynamics
in general. For example Halofit corrections to the matter power spectrum
at small scales are calibrated from $\Lambda$CDM numerical simulations
and are not necessarily meaningful in models departing from concordance.

\subsection{Available Models\label{sub:Models}}

In this first public release of \texttt{hi\_class}, we have
only enabled a limited number of models. For the initial roll-out, we have
focused on the EFT-like parameterisations, leaving models specified
through full covariant actions for the next release. Moreover, the
current implementation is limited to flat spatial sections for the
cosmological background, $\Omega_{k}=0$.

The user chooses the model using the initialization file \texttt{hi\_class.ini}.
This or a similar file is read in by the input module, which sets
up the relevant variables and performs the appropriate tests.\footnote{See the \texttt{hi\_class.ini} file included with the code for a detailed
description of the available user options. The \texttt{hi\_class}
wiki contains extended and up-to-date  information (\url{https://github.com/miguelzuma/hi_class_public/wiki}).} The user needs to specify the DE density and pressure and the four
$\alpha$ functions. In general, they can be functions of any of the
background quantities (e.g.~$H,a,\tau$, etc.).

For the background energy density for the Horndeski scalar we have
enabled two choices in this first version of \texttt{hi\_class }(see
table \ref{tab:Parameterized-expansion}): 
\begin{enumerate}
\item \texttt{lcdm}: the DE has constant energy density determined by the
parameter $\Omega_{{\rm DE},0}$ derived from the closure relation
(\ref{eq:Closure}).
\item \texttt{wowa}: evolving equation of state with parameters $\left(\Omega_{\text{DE},0},w_{0},\,w_{a}\right)$
\cite{Chevallier:2000qy,Linder:2002et} with the constraint that the
density fraction today satisfies the closure relation (\ref{eq:Closure}).
\end{enumerate}
These are sufficient to fully evolve the cosmological background.
The user must then specify a parameterisation for the $\alpha$ functions
needed to calculate the evolution of perturbations on this background.
At this juncture we are enabling four parameterisations for the $\alpha$
functions (see table \ref{tab:Parameterized-models}):
\begin{enumerate}
\item \texttt{propto\_omega}: the $\alpha$ functions are proportional to
the density fraction of dark energy, $\Omega_{\text{DE}}$
\item \texttt{propto\_scale}: the $\alpha$ functions are proportional to
the scale factor 
\item \texttt{planck\_linear}: k-\emph{essence} conformally coupled to gravity
studied by Planck \cite{Ade:2015rim}
\item \texttt{planck\_exponential}: alternative k-\emph{essence} conformally
coupled to gravity studied by Planck \cite{Ade:2015rim}
\end{enumerate}
All these parameterisations have $\alpha_{i}\rightarrow0$ at early
times, consistent with the expectation that gravity would not be modified
during eras when the dark energy does not provide a significant contribution
to the energy density of the Universe. 

We have also implemented two models which were studied by the Planck
team \cite{Ade:2015rim}, \texttt{planck\_linear} and \texttt{planck\_exponential},
to provide a possibility of easy comparison with published constraints.
Both these models are specified through a single function of time
$\Omega\left(\tau\right)$ affecting three of the $\alpha$ functions:

\begin{equation}
\begin{cases}
\alpha_{\textrm{K}}= & \frac{3\left(\rho_{\textrm{DE}}+p_{\textrm{DE}}\right)}{H^{2}}+\frac{3\Omega\left(\rho_{\textrm{m}}+p_{\textrm{m}}\right)}{H^{2}\left(1+\Omega\right)}-\frac{\Omega^{\prime\prime}-2aH\Omega^{\prime}}{a^{2}H^{2}\left(1+\Omega\right)}\\
\alpha_{\textrm{M}}= & \frac{\Omega^{\prime}}{aH\left(1+\Omega\right)}\\
\alpha_{\textrm{B}}= & -\alpha_{\textrm{M}}\\
\alpha_{\textrm{T}}= & 0\,.
\end{cases}\label{eq:planck_parametrization}
\end{equation}
The difference between the ``linear'' and the ``exponential''
models is just the parameterisation for $\Omega(\tau)$ (see table
\ref{tab:Parameterized-models}).

\begin{table}
\begin{centering}
\subfloat[Parameterisations for dark energy density enabled in the first version
of \texttt{hi\_class}. \label{tab:Parameterized-expansion}]{\begin{centering}
\begin{tabular}{lccccl}
\toprule 
\multicolumn{6}{l}{\textsf{\textbf{Parameterisations of Background }}}\tabularnewline
\texttt{expansion\_model\_smg} & \multicolumn{4}{l}{\textsf{\textbf{\footnotesize{}Expansion}}} & \textsf{\textbf{\footnotesize{}Notes}}\tabularnewline
\midrule
\midrule 
\texttt{lcdm} & \multicolumn{4}{l}{$\rho_{\textrm{DE}}=\Omega_{{\rm DE},0}H_{0}^{2}$} & CC-like\tabularnewline
\midrule 
\texttt{w0wa} & \multicolumn{4}{l}{$\rho_{\textrm{DE}}=\Omega_{{\rm DE},0}H_{0}^{2}a^{-3\left(1+w_{0}+w_{a}\right)}e^{3w_{a}(a-1)}$} & Similar to \texttt{omega\_fld}\tabularnewline
\bottomrule
\end{tabular}
\par\end{centering}

}
\par\end{centering}

\begin{centering}
\subfloat[Parameterisations of the $\alpha$ functions enabled in the first
version of \texttt{hi\_class},\label{tab:Parameterized-models}]{\begin{centering}
\begin{tabular}{lccccc}
\toprule 
\multicolumn{6}{l}{\textsf{\textbf{Parameterisations of Gravity}}}\tabularnewline
\texttt{gravity\_model\_smg} & \multicolumn{4}{c}{\quad{}$\alpha_{\text{K}}$\quad{}\quad{}$\alpha_{\text{B}}$\quad{}\quad{}$\alpha_{\text{M}}$\quad{}\quad{}$\alpha_{\text{T}}$\quad{}} & \textsf{\textbf{\footnotesize{}Additional Params.}}\tabularnewline
\midrule
\midrule 
\texttt{propto\_omega} & \multicolumn{4}{c}{$\alpha_{i}=\hat{\alpha}_{i}\Omega_{\text{DE}}(\tau)$} & $M_{*,{\rm ini}}^{2}$\tabularnewline
\midrule 
\texttt{propto\_scale} & \multicolumn{4}{c}{$\alpha_{i}=\hat{\alpha}_{i}a(\tau)$ } & $M_{*,{\rm ini}}^{2}$\tabularnewline
\midrule 
\texttt{planck\_linear} & \multicolumn{4}{c}{Eq. (\ref{eq:planck_parametrization}) with $\Omega=\Omega_{0}a\left(\tau\right)$} & $M_{*,{\rm ini}}^{2}$\tabularnewline
\midrule 
\texttt{planck\_exponential ~~} & \multicolumn{4}{c}{Eq. (\ref{eq:planck_parametrization}) with $\Omega=\exp\left[\frac{\alpha_{\textrm{M}0}}{\beta}a^{\beta}\left(\tau\right)\right]-1$} & $M_{*,{\rm ini}}^{2}$\tabularnewline
\bottomrule
\end{tabular}
\par\end{centering}

}
\par\end{centering}

\centering{}\caption{Summary of models enabled in the first version of \texttt{hi\_class}.
For further details read the \texttt{hi\_class.ini} file or visit \protect\url{https://github.com/miguelzuma/hi_class_public/wiki}. \label{tab:Summary-of-models}}
\end{table}

\subsection{Examples\label{sub:Plots}}

\begin{figure}
\begin{centering}
\includegraphics[width=0.49\textwidth]{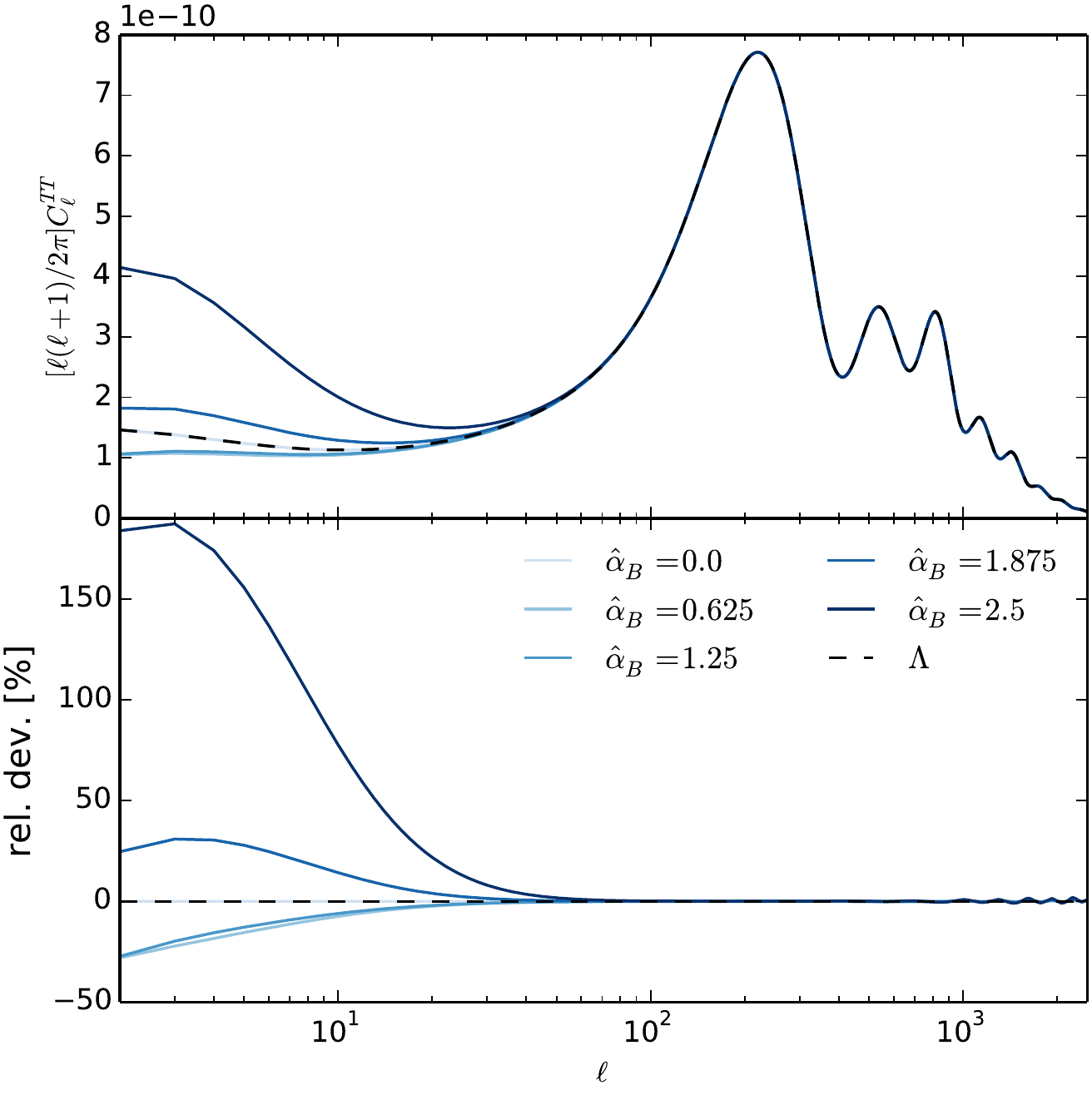} \includegraphics[width=0.5\textwidth]{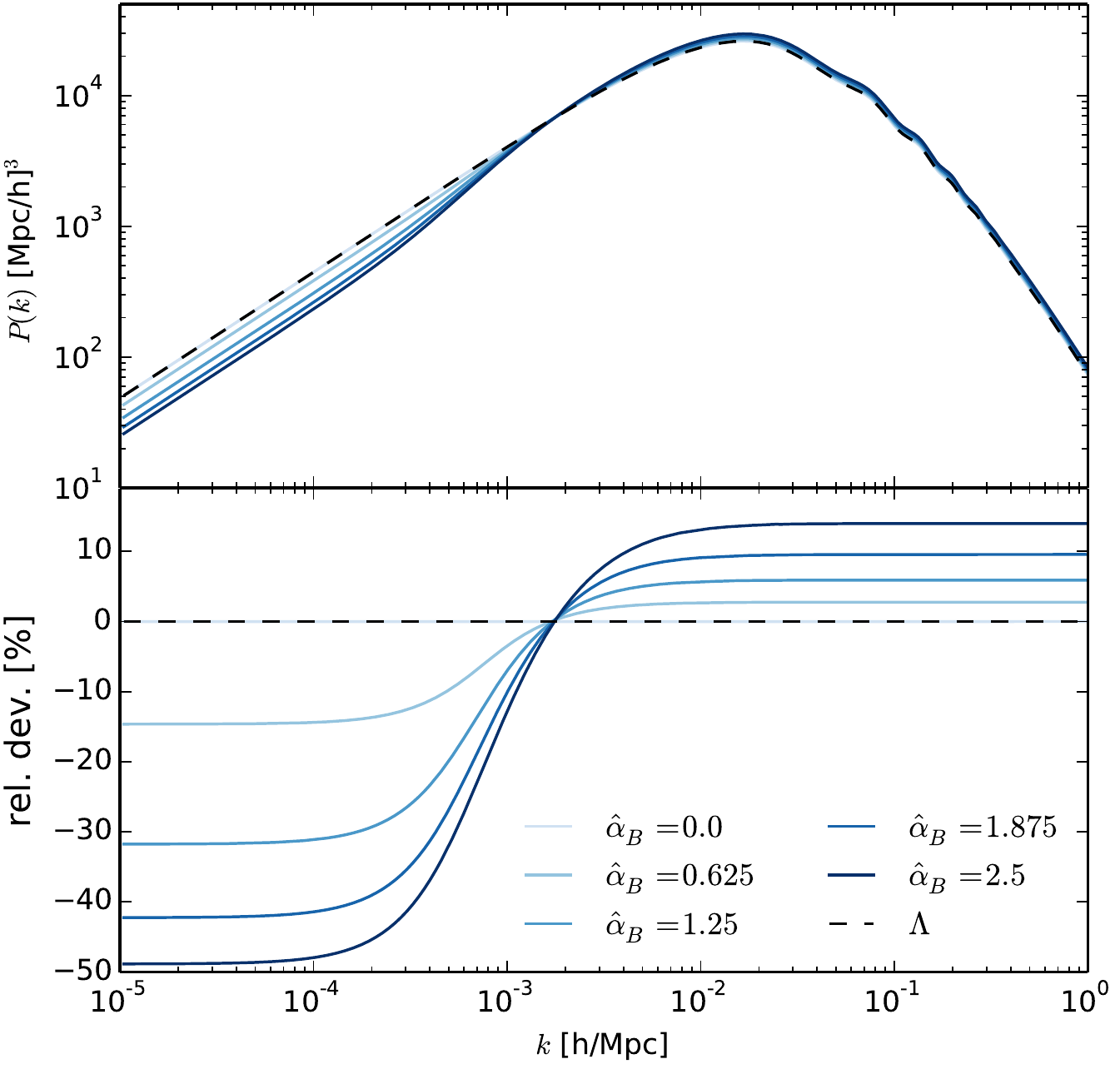}
\par\end{centering}

\caption{Effect of the braiding on CMB temperature power spectra (left panel)
and the matter power spectrum at $z=0$ (right panel). In the bottom
panels we show the relative difference between the DE/MG models considered
and a fiducial $\Lambda$CDM. In each model, we fix all the standard
cosmological parameters to some fiducial value. We use the \texttt{lcdm}
parametrization for the background expansion history (i.e.~$w_{\textrm{DE}}=-1$)
and \texttt{propto\_omega} for the evolution of the alphas. We fix
$\hat{\alpha}_{\textrm{K}}=1$, $\hat{\alpha}_{\textrm{M}}=\hat{\alpha}_{\textrm{T}}=0$
and we vary $\hat{\alpha}_{\textrm{B}}$ in the stable region.\label{fig:alpha_B}}
\end{figure}

\begin{figure}
\bigskip{}

\begin{centering}
\includegraphics[width=0.49\textwidth]{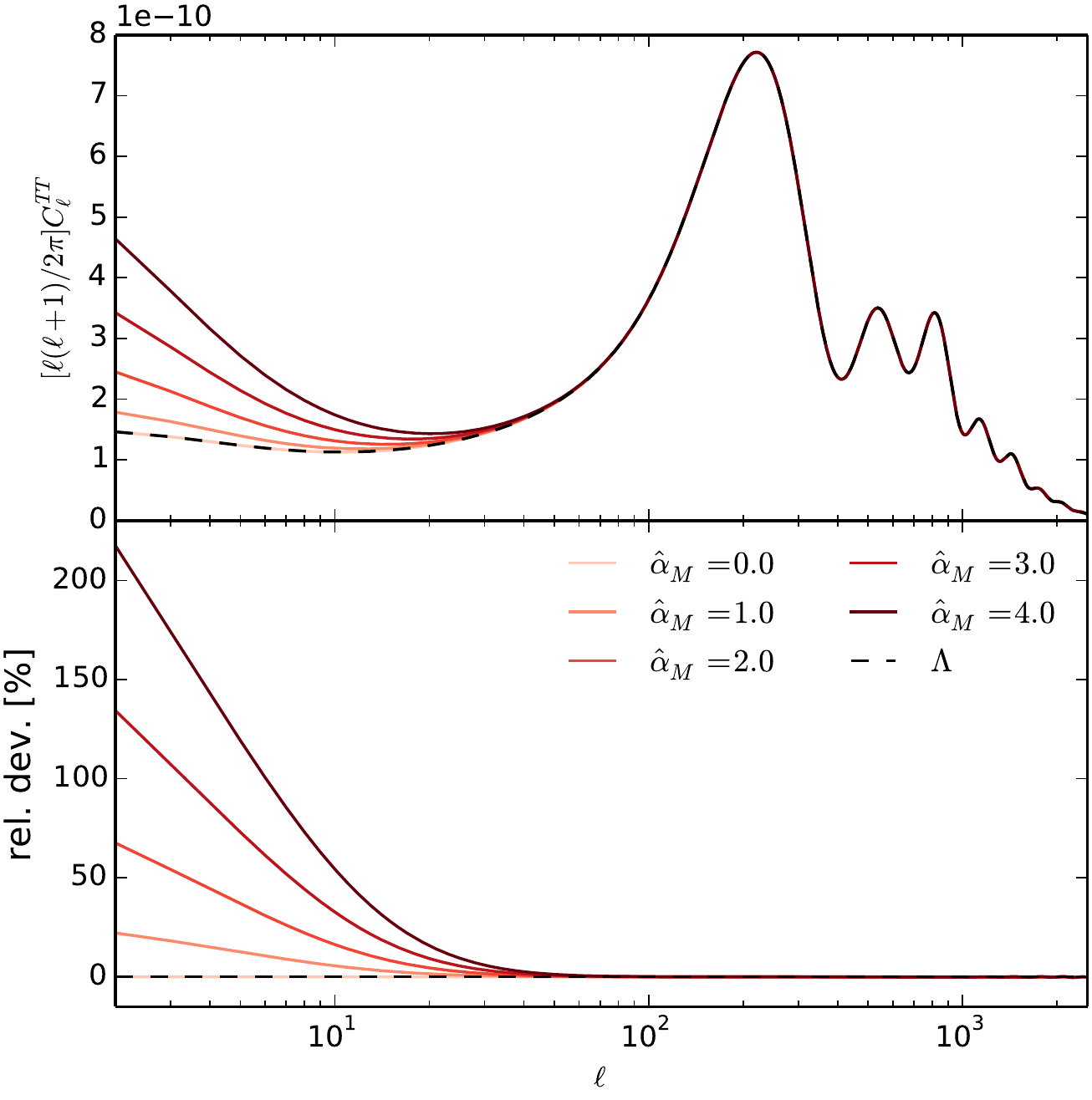} \includegraphics[width=0.5\textwidth]{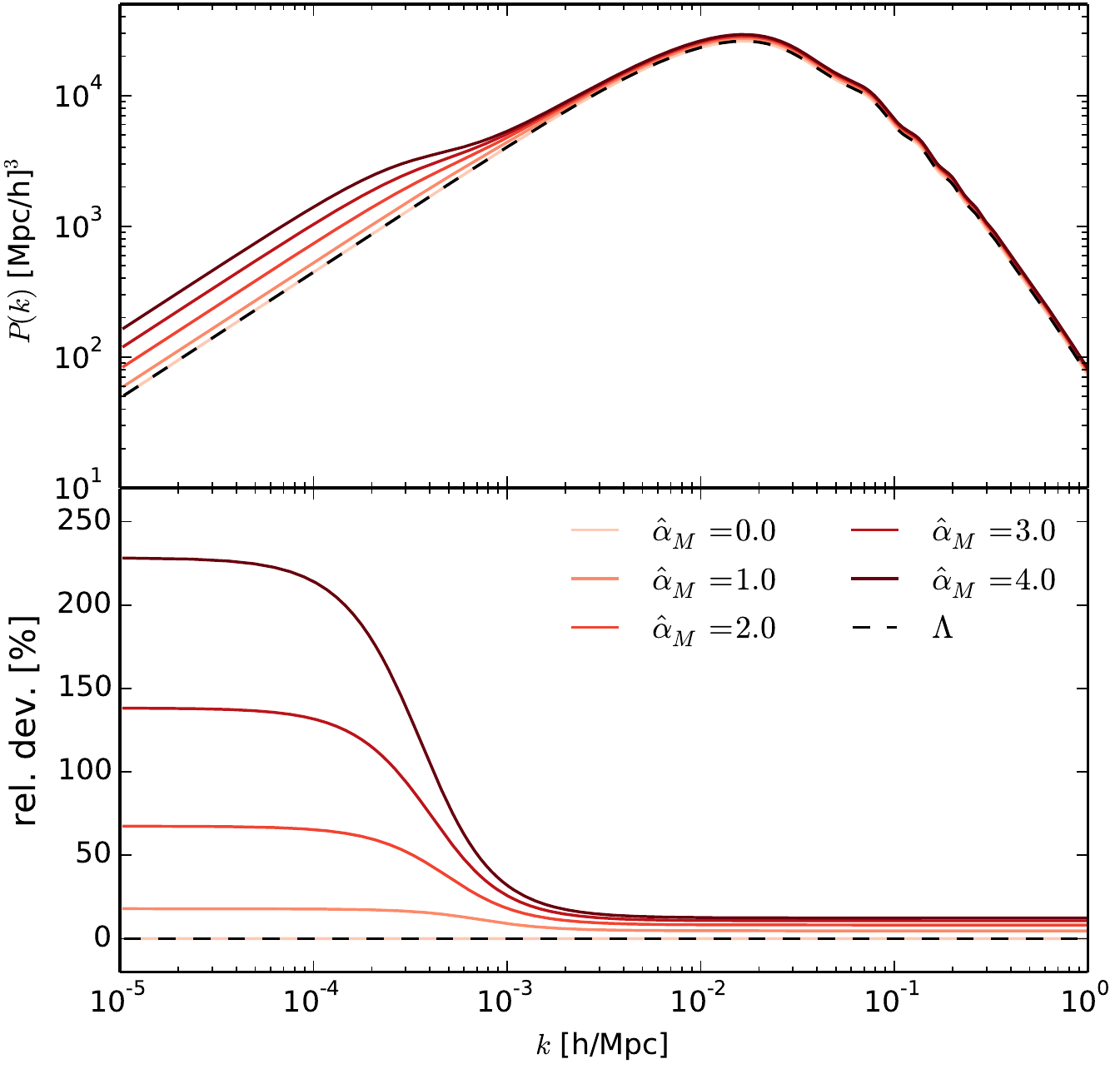}
\par\end{centering}

\caption{Same as Figure \ref{fig:alpha_B} but for the running of the Planck
mass. We fix $\hat{\alpha}_{\textrm{K}}=1$, $\hat{\alpha}_{\textrm{B}}=\hat{\alpha}_{\textrm{T}}=0$
and we vary $\hat{\alpha}_{\textrm{M}}$ in the stable region, with
$M_{*,0}^{2}=1$ initially.\label{fig:alpha_M}}
\end{figure}

\begin{figure}
\bigskip{}

\begin{centering}
\includegraphics[width=0.49\textwidth]{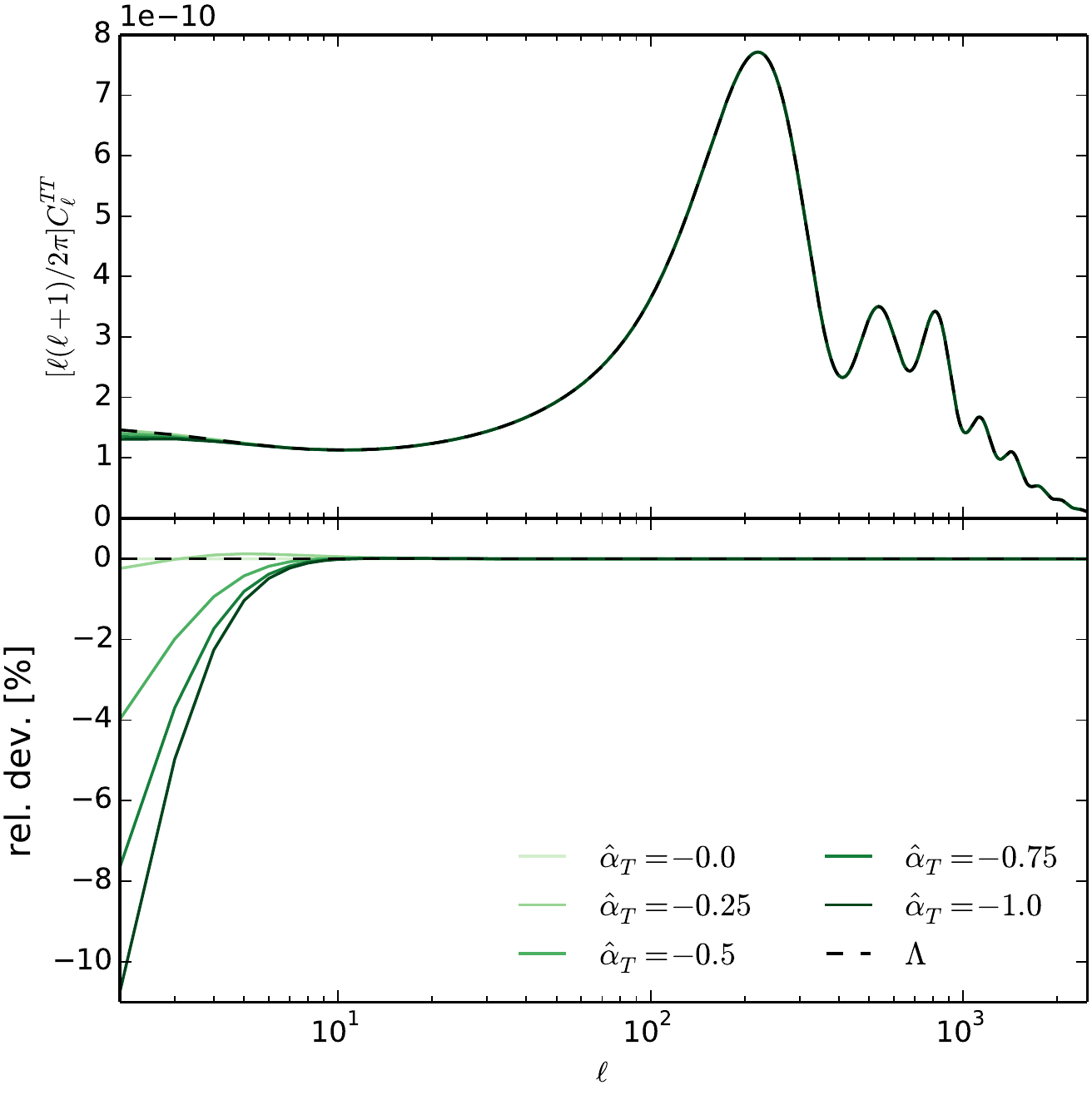} \includegraphics[width=0.5\textwidth]{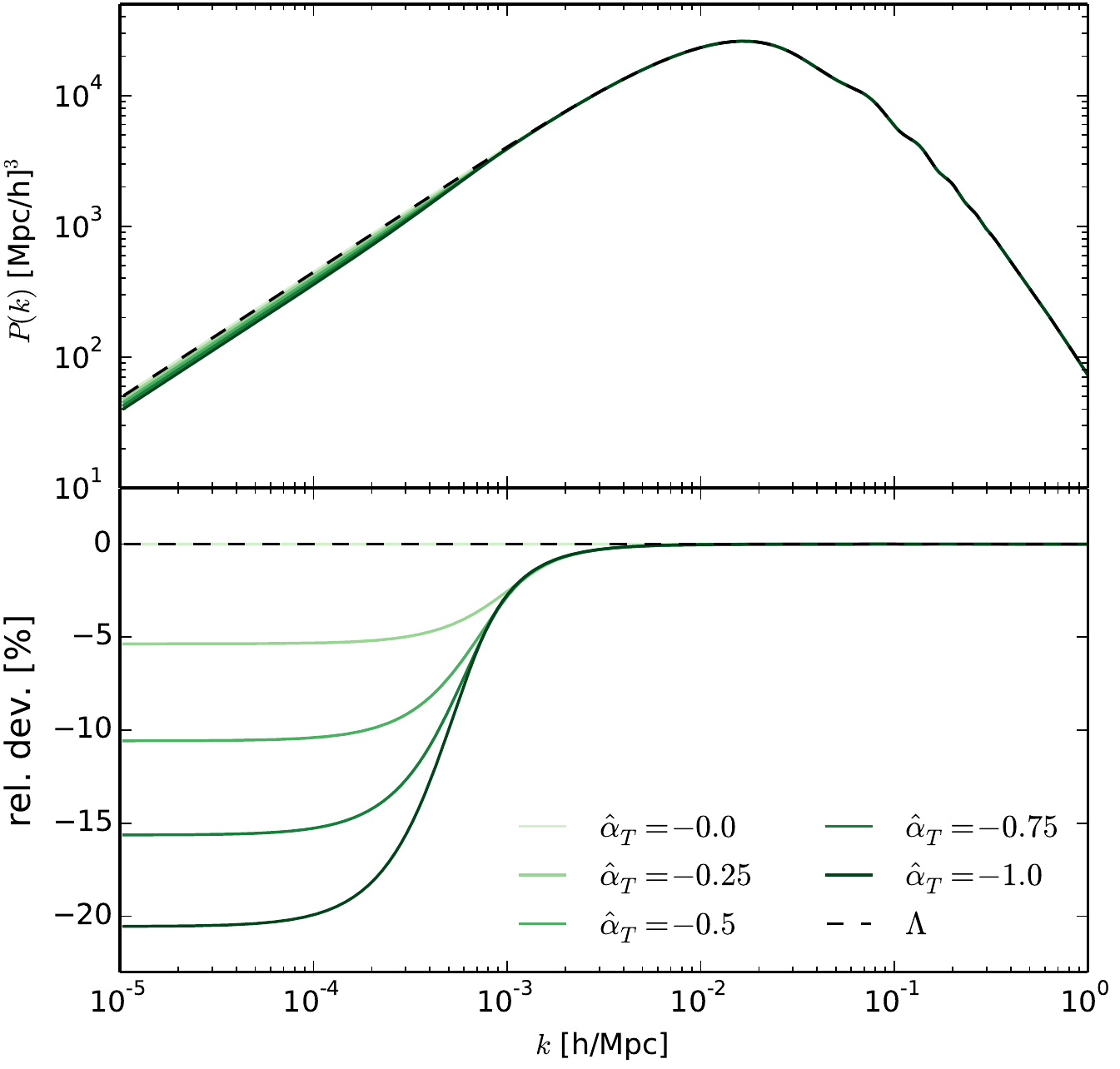}
\par\end{centering}

\caption{Same as Figure \ref{fig:alpha_B} but for the tensor speed excess.
We fix $\hat{\alpha}_{\textrm{K}}=1$, $\hat{\alpha}_{\textrm{B}}=\hat{\alpha}_{\textrm{M}}=0$
and we vary $\hat{\alpha}_{\textrm{T}}$ in the stable region.\label{fig:alpha_T}}
\end{figure}

In Figs.~\ref{fig:alpha_B}, \ref{fig:alpha_M} and \ref{fig:alpha_T}
we show some illustrative plots of the CMB temperature power spectrum
(left panels) and the matter power spectrum calculated at redshift
$z=0$ (right panels). In all these plots we have fixed the standard
cosmological parameters to Planck-best-fit values \cite{Ade:2015xua}
and we vary only the DE/MG parameters. They have been created using
the \texttt{lcdm} parametrization for the background expansion history
and \texttt{propto\_omega} for the time evolution of the $\alpha$
functions. Constraints using this parametrization and recent data
have been obtained in ref.~\cite{Bellini:2015xja}. Here the authors
also show that variations of the kineticity have a negligible effect
on this background. We thus fix $\hat{\alpha}_{\text{K}}=1$ and explore
the remaining $\alpha$ functions.

In particular, in Fig.~\ref{fig:alpha_B} we fix $\hat{\alpha}_{\textrm{K}}=1$,
$\hat{\alpha}_{\textrm{M}}=\hat{\alpha}_{\textrm{T}}=0$ and we vary
$\hat{\alpha}_{\textrm{B}}$. With these restrictions, negative values
of $\hat{\alpha}_{\textrm{B}}$ are not allowed by the stability conditions,
Eqs.~(\ref{eq:scalar_ghost}-\ref{eq:tensor_grad}). It is possible
to note that the relative difference between DE/MG models and the
fiducial $\Lambda$CDM is large at small scales, i.e.~low-$\ell$'s
and small $k$. In particular, the amplitude of the matter power spectrum
is suppressed on these scales. However, $\hat{\alpha}_{\textrm{B}}$
has also a non-negligible effect at large $k$, i.e.~$k\gtrsim10^{-2}$,
where we can see the opposite effect, i.e.~the amplitude of the matter
power spectrum is enhanced as $\hat{\alpha}_{\textrm{B}}$ increases.
This results from the scalar field communicating an extra force and
increasing the attractive interaction between perturbations at small
scales \cite{Bellini:2014fua}. Note that increasing $\hat{\alpha}_{\text{B}}$
pushes the low-$\ell$ CMB to lower values, but then becomes rapidly
larger than in the standard case. This is because the braiding tends
reduce the Integrated Sachs-Wolfe (ISW) and even make it negative
\cite{Renk:2016olm}. The initial decrease in the CMB spectrum is
due to this initial reduction, while the subsequent increase for $\hat{\alpha}_{B}\gtrsim1.5$
is due to the positive definite $(\text{ISW})^{2}$ contribution.

In Fig.~\ref{fig:alpha_M} we fix $\hat{\alpha}_{\textrm{K}}=1$,
$\hat{\alpha}_{\textrm{B}}=\hat{\alpha}_{\textrm{T}}=0$ and we vary
$\hat{\alpha}_{\textrm{M}}$, setting $M_{*,ini}^{2}=1$ initially.
In this case negative values of $\hat{\alpha}_{\textrm{M}}$ lead
to gradient instabilities. Then, assuming positive values of the Planck
mass run-rate it is possible to note that the CMB temperature spectra
amplitude are enhanced at low $\ell$'s due to the ISW effect, while
for large values of $\ell$ there are not substantial differences.
An analogous result can be found looking at the matter power spectrum
plots, where at large scales the relative differences between DE/MG
models and $\Lambda$CDM appear maximised.

In Fig.~\ref{fig:alpha_T} we fix $\hat{\alpha}_{\textrm{K}}=1$,
$\hat{\alpha}_{\textrm{B}}=\hat{\alpha}_{\textrm{M}}=0$ and we vary
$\hat{\alpha}_{\textrm{T}}$. Positive values of $\hat{\alpha}_{\textrm{T}}$
are unstable with this choice of the other parameters, and thus we have explored only the negative region. The effect of the tensor speed excess on the CMB temperature
spectrum and the matter power spectrum is smaller w.r.t.~the effects
seen in Figs.~\ref{fig:alpha_B} and \ref{fig:alpha_M} at all scales
even for extreme values of $\hat{\alpha}_{\textrm{T}}$. This indicates
that $\hat{\alpha}_{\textrm{T}}$ is expected to contribute less than
the other parameters on CMB and LSS observables. In particular, $\hat{\alpha}_{\text{T}}$
by itself has no effect whatsoever on small scales.

We should also note that we have generated the plots for $w_{\text{DE}}=-1$.
In this limit, many terms in the perturbation equations cancel and
therefore the modifications to observables are suppressed. Adding
in a background evolution that departs from $\Lambda$CDM can produce
a much richer phenomenology. Similarly, considering simultaneously
non-zero values of several $\alpha$ functions can enhance the observed
effects and produce new ones, because of products of $\alpha$ functions
appearing in the equations of motion. We emphasize that the spectra
are the result of solving a complex system of differential equations
(plus performing line-of-sight integrals in the case of the CMB) and
therefore it is not to be expected for the deviations to be proportional
to the coefficients $\hat{\alpha}_{i}$. 
As a result, the posteriors
for the $\hat{\alpha_{i}}$ parameters are not expected to be Gaussian
and are thus sensitive to the choice of the fiducial model.

\section{Summary and Outlook}

This paper accompanies the launch of the first public version
of \texttt{hi\_class}, an implementation of Horndeski's theory in
the Cosmic Linear Anisotropy Solving System. This initial version allows
the user to freely specify parameterisations of the expansion history,
as well as the functions that characterise the evolution of the perturbations
at the linear level. The parameterised approach can be used to test
gravity in a largely model-independent manner by directly modifying
the fundamental properties of the gravitational degrees of freedom
and calculating the subsequent impact on cosmological observables.
Although technically more involved, this way of working neatly generalises
the widely used parameterisations of the equation of state $w_{\textrm{DE}}$:
just as detecting $w_{\textrm{DE}}\neq-1$ on the Hubble diagram would
be a clear signature of physics beyond the cosmological constant,
any significant data preference for $\alpha_{i}\neq0$ would constitute
model-independent evidence for dynamics beyond Einstein gravity.

The formulation of modified gravity as implemented in \texttt{hi\_class} ensures that whatever the choice of parameters is made, there exists a model in the Horndeski class of scalar-tensor theories which would have exactly such a phenomenology at linear level in perturbations. The phenomenology of \emph{any} particular scalar-tensor model of gravity or dark energy (such as quintessence, $f(R)$ or galileons) can be obtained by choosing particular $\alpha$ functions and background. In this way, \texttt{hi\_class}, should be thought of as a unified code to test all such modifications of gravity for which separate codes exist currently. 

The current version of \texttt{hi\_class} represents just the first
step towards a reliable, fast and flexible tool to address modifications
of gravity on cosmological scales in a self-consistent manner. Further extensions 
and improvements to \texttt{hi\_class} will be released publicly in the near
future and which will include \emph{inter alia} approximation schemes to
optimise the computational performance, automatic predictions for
full covariant action-based models and also expand further the range
of theories available to test.

The parameterised approach as implemented in the public \texttt{hi\_class}
currently relies neither on a specific choice for the action (\ref{eq:action})
nor on the initial conditions for the scalar field at the cosmological
background, but rather is a very general method to test for evidence
of non-GR effects in scalar tensor theories. This sort of study should be complemented by testing
models arising from full covariant actions and placing limits on their
parameters. One of the main features of the full version of \texttt{hi\_class}
currently in development is to allow the user to directly specify
the Horndeski functions $G_{i}$ in (\ref{eq:L2}-\ref{eq:L5}), integrate
the background scalar equation given some intial conditions and automatically
generate predictions in terms of the parameters that define the model.
Starting with a covariant action-based model offers important advantages
compared to the EFT approach:
\begin{enumerate}
\item \emph{Background expansion and Self-Consistency.} In a particular model, the background
and perturbations are related and controlled by the same parameters.
In many Horndeski theories the background departs considerably from
the $\Lambda$CDM or $w$CDM, leading to significantly stronger constraints
from the outset (e.g.~in galileons \cite{Barreira:2014jha}). Moreover,
certain time dependences and values of the $\alpha$ functions might
require very fine tuned parameters, even when very general Lagrangians
are considered (e.g.~as in quintessence \cite{Marsh:2014xoa}, also
see \cite{Linder:2015rcz}).
\item \emph{Cosmology beyond linear order.} Extending the EFT approach to
general models beyond linear order introduces a severely expanded
set of functions at each order and rapidly spoils the economy possible
in the linear description. When considering the Horndeski class of
theories, this issue does not arise, since the structure of the theory
is fixed and further relations are available in terms of
the $G_{i}$ functions \cite{Bellini:2015wfa}. In principle, the
results of the linear predictions of \texttt{hi\_class} can also be
connected to the limit of very non-linear structure formation, e.g.~through
N-body simulations for specific models (e.g.~galileons \cite{Li:2013tda})
including understanding the effects of the particular screening built
into the model.
\item \emph{Relation to other regimes.} When working with a specified model,
it becomes possible to link constraints from cosmology to tests of
gravity in any other regime. This includes not only Solar-System tests,
but also generation of gravitational waves, compact objects and other
astrophysical tests \cite{Berti:2015itd} as well as, potentially,
laboratory and collider experiments to connect tests of gravity on
a broad range of scales \cite{Baker:2014zba}. Moreover, starting
with a specific action allows one to understand the theoretical limits
on the model, e.g.~to compute the strong-coupling scale and to understand
the range of validity of this description. Finally, it might even
be possible to find connections with scenarios in fundamental physics
such as extra dimensions and quantum gravity. 
\end{enumerate}
For these reasons, one should continue to test models defined through
a covariant action which would connect predictions in various regimes.
Nonetheless, the parameterised EFT allows for an exploration of the
data for effects arising in a possibly wider set of models which remain
consistent with the underlying symmetries assumed in cosmology but
do not fit into the Horndeski scheme. The ultimate design goal of
\texttt{hi\_class} is to achieve full generality as a tool to test
gravity with cosmological data. To this end future releases of \texttt{hi\_class}
will transcend the scope of Horndeski's theory and include the full
EFT of DE and its generalizations \cite{Lagos:2016wyv,Langlois:2017mxy}. 

\texttt{hi\_class} solves the full dynamical equations consistently
across all cosmic epochs and scales, relying only on the validity
of the linear description, but not making use of the so-called quasi-static
approximation. A future design goal is to implement the QS approximation
in \texttt{hi\_class} in a controlled manner, not as a means to simplify
the results, but rather as a tool to speed up the computations on
scales and epochs in which these dynamical effects are below the desired
precision. Future releases will also extend the choice of initial
conditions for the scalar field, allowing for an exploration of models
that lie beyond the usual scope of late time acceleration, including
those featuring modifications of gravity in early epochs and connecting
directly with inflationary physics.

In parallel, we are enhancing the code to compute observables beyond
the linear regime, including higher-order effects \cite{Bellini:2015oua}
and improved perturbation-theory schemes suitably adapted to theories
of gravity beyond Einstein's general relativity. These improvements
will aim at making \texttt{hi\_class} into modern tool able to match
the requirements of the upcoming era of precision cosmology, helping
the future generation of large LSS surveys to unveil the mysteries
of the dark universe and shed light on the nature of gravity.

\paragraph*{Acknowledgements: }

We are especially grateful to Thomas Tram for patiently answering
many questions pertaining to the CLASS code, Alexandre Barreira for
providing us with the output to test the code, and David Alonso, Nicola Bellomo,
Antonio J.~Cuesta, Carlos Garcia-Garcia, Francesco Montanari, Janina Renk, Angelo Ricciardone
and Alessio Spurio Mancini, for using preliminary versions of \texttt{hi\_class}
and providing us with valuable suggestions leading to many improvements.
We are also thankful to Jose María Ezquiaga, Eric Linder, Francesco Montanari and
Janina Renk for comments on the first version as well as to Luca Amendola,
Guillermo Ballesteros, Dario Bettoni, Alicia Bueno-Belloso, Santiago Casas,  Juan García-Bellido, Cristiano Germani, Jérôme Gleyzes, Kurt Hinterbichler, Gregory Horndeski, Bin Hu, Luisa Jaime, Raul Jimenez, Valeria Pettorino, Marco Raveri, Laura Taddei, Lorenzo Tamellini, Licia Verde and Filippo
Vernizzi for useful conversations during the completion of this work.
The \texttt{hi\_class }logo was designed by Marcos Vázquez Pingarrón
(\url{www.mywonderland.es}).

This project was initiated and substantially carried out while E.B.~and
M.Z.~benefited from being members of the Institut für Theoretische
Physik of the University of Heidelberg. The work of E.B. is partially supported
by the European Research Council under the European Community’s Seventh
Framework Programme FP7-IDEAS-Phys.LSS 240117  and the ERC H2020 693024 GravityLS project as well as the Spanish
MINECO through projects AYA2014-58747-P and MDM-2014- 0369 of ICCUB
(Unidad de Excelencia María de Maeztu). 
I.S. is supported by the European
Regional Development Fund and the Czech Ministry of Education, Youth
and Sports (MŠMT) (Project CoGraDS — CZ.02.1.01/0.0/0.0/15\textbackslash{}\_003/0000437).
While in Heidelberg M.Z. acknowledges support from DFG through the TRR33
project ``The Dark Universe''. PGF is supported by ERC H2020 693024 GravityLS project, the Beecroft Trust and STFC.

\appendix

\section{Equations and notation\label{sec:Equations-and-notation}}

In this section, we list all the equations used in \texttt{hi\_class},
both at the background and at the perturbation level. It is important
to note that the notation of CLASS is slightly different than the
notation of ref.~\cite{Bellini:2014fua}. The matter densities are
redefined in such a way that the Friedmann constraint is written as
$H^{2}=\underset{i}{\sum}\rho_{i}$. In addition, it should be noted
that all the variables are expressed in conformal time $\tau$ (a
prime denotes derivative w.r.t.~conformal time), while the Hubble
parameter is the physical one, i.e.~$a^{\prime}=a^{2}H$. 

The line element in synchronous gauge up to first order in perturbation
theory reads \cite{Ma:1995ey}
\begin{equation}
ds^{2}=a^{2}\left[ -d\tau^{2}+\left(\delta_{ij} + \tilde{h}_{ij}\right)dx^{i}dx^{j}\right] \,,\label{eq:metric}
\end{equation}
where in Fourier space
\begin{equation}
 \tilde{h}_{ij}\left(\tau, \vec{k}\right)=  \hat{k}_i \hat{k}_j h + 6\left(\hat{k}_i \hat{k}_j - \frac{1}{3}\delta_{ij}\right) \eta + h_{ij}\,.
\end{equation}
Here $h$ and $\eta$ are scalar perturbations and $h_{ij}$ is the tensor perturbation. Sometimes in the code it is used the perturbation $\xi\left(\tau,\vec{k}\right)$, which is related to $h$ and $\eta$ through\footnote{Note that in ref.~\cite{Ma:1995ey} our $\xi$ is named $\alpha$. We modified the notation to avoid confusion with the $\alpha$ functions
of modified gravity.}
\begin{equation}
\xi=\frac{h^{\prime}+6\eta^{\prime}}{2k^{2}}\,.
\end{equation}

\subsection{Background\label{sec:AppBack}}

\texttt{hi\_class} makes the choice of units such that the Friedmann
equations read
\begin{align}
H^{2}= & \rho_{\textrm{m}}+\rho_{\textrm{DE}}\\
H^{\prime}= & -\frac{3}{2}a\left(\rho_{\textrm{m}}+p_{\textrm{m}}+\rho_{\textrm{DE}}+p_{\textrm{DE}}\right)\,,
\end{align}
where $\rho_{\textrm{m}}$ and $p_{\textrm{m}}$ are the energy density
and pressure of the total matter content of the universe (except dark
energy), while $\rho_{\textrm{DE}}$ and $p_{\textrm{DE}}$ are the
energy density and pressure of the scalar field
\begin{align}
\mathcal{\rho_{\text{DE}}}\equiv & -\frac{1}{3}G_{2}+\frac{2}{3}X\left(G_{2X}-G_{3\phi}\right)-\frac{2H^{3}\phi^{\prime}X}{3a}\left(7G_{5X}+4XG_{5XX}\right)\label{eq:rhoDE-1}\\
 & +H^{2}\left[1-\left(1-\alpha_{\textrm{B}}\right)M_{*}^{2}-4X\left(G_{4X}-G_{5\phi}\right)-4X^{2}\left(2G_{4XX}-G_{5\phi X}\right)\right]\nonumber \\
p_{\text{DE}}\equiv & \frac{1}{3}G_{2}-\frac{2}{3}X\left(G_{3\phi}-2G_{4\phi\phi}\right)+\frac{4H\phi^{\prime}}{3a}\left(G_{4\phi}-2XG_{4\phi X}+XG_{5\phi\phi}\right)-\frac{\left(\phi^{\prime\prime}-aH\phi^{\prime}\right)}{3\phi^{\prime}a}HM_{*}^{2}\alpha_{\textrm{B}}\label{eq:pDE-1}\\
 & -\frac{4}{3}H^{2}X^{2}G_{5\phi X}-\left(H^{2}+\frac{2H^{\prime}}{3a}\right)\left(1-M_{*}^{2}\right)+\frac{2H^{3}\phi^{\prime}XG_{5X}}{3a}\,,\nonumber 
\end{align}
(note that both eqs (\ref{eq:rhoDE-1}) and (\ref{eq:pDE-1}) contain $\alpha_{\text{B}}$, defined in eq. (\ref{eq:aB})). 

The manner in which we treat $M_{*}^{2}$ here is different than in ref.~\cite{Bellini:2014fua}. This allows us to ensure the usual conservation equations, both for
the scalar field and the matter, 
\begin{align}
\rho_{\textrm{DE}}^{\prime} & +3aH\left(\rho_{\textrm{DE}}+p_{\textrm{DE}}\right)=0\,,\\
\rho_{\text{m}}' & +3aH(\rho_{\text{m}}+p_{\text{m}})=0\ .\nonumber 
\end{align}
which replicates the meaning of the usual parameterisation of $w_{\text{DE}}$
as a description of the expansion history and thus is much simpler
to communicate. On the other hand, we note that an observer comoving
with the cosmological background in the past would not be aware of
the evolving strength of gravity. Thus if a local measurement of the
energy density and pressure were possible and performed, the observer
would rather see the quantities with \textasciitilde{},
\begin{align}
\tilde{\rho}_{\textrm{m}}= & \frac{3\rho_{\textrm{m}}}{M_{*}^{2}}\,,\\
\tilde{p}_{\textrm{m}}= & \frac{3p_{\textrm{m}}}{M_{*}^{2}}\,,\\
\tilde{\mathcal{E}}= & \frac{3\rho_{\textrm{DE}}}{M_{*}^{2}}+3H^{2}\frac{\left(M_{*}^{2}-1\right)}{M_{*}^{2}}\,,\\
\tilde{\mathcal{P}}= & \frac{3p_{\textrm{DE}}}{M_{*}^{2}}-\left(3H^{2}+\frac{2H^{\prime}}{a}\right)\frac{\left(M_{*}^{2}-1\right)}{M_{*}^{2}}\,,
\end{align}
as defined in \cite{Bellini:2014fua} and thus would have measured
a different equation of state.

\subsection{Linear perturbations in Synchronous Gauge\label{sub:Synchronous_perturbations}}

At linear order in perturbation theory, CLASS has the possibility to
solve the evolution of the perturbations in both Newtonian and synchronous
gauges. The present version of \texttt{hi\_class} implements only
the synchronous gauge option. 

First, the Horndeski class of models modifies the propagation of gravitational
waves $h_{ij}$ (tensors). This has been taken into account in \texttt{hi\_class}
through the dynamical equation 

\begin{equation}
h_{ij}^{\prime\prime}+\left(2+\alpha_{\textrm{M}}\right)aHh_{ij}^{\prime}+\left(1+\alpha_{\textrm{T}}\right)k^{2}h_{ij}=\frac{\sigma_{\textrm{m}ij}}{M_{*}^{2}}\,,\label{eq:tesors}
\end{equation}
where $\sigma_{\textrm{m}ij}$ represents the source for the tensor
modes arising from the matter.

The scalar sector is described through the perturbations of the metric
tensor $\left(\eta,\,h,\,\xi\right)$ (see eq.~(\ref{eq:metric})),
the scalar field perturbation $V_{X}$ (defined in eq.~(\ref{eq:VXdef})),
and the perturbations of the matter density, velocity, pressure and
anisotropic stress, $\left(\delta\rho_{\textrm{m}},\,\theta_{\textrm{m}},\,\delta p_{\textrm{m}},\,\sigma_{\textrm{m}}\right)$
respectively. The Einstein equations are
\begin{itemize}
\item Einstein (0,0)
\begin{align}
h^{\prime}= & \frac{4k^{2}\eta}{aH\left(2-\alpha_{\textrm{B}}\right)}+\frac{6a\delta\rho_{\textrm{m}}}{HM_{*}^{2}\left(2-\alpha_{\textrm{B}}\right)}-2aH\left(\frac{\alpha_{\textrm{K}}+3\alpha_{\textrm{B}}}{2-\alpha_{\textrm{B}}}\right)V_{X}^{\prime}\nonumber \\
 & -2\left[3aH^{\prime}+\left(\frac{\alpha_{\textrm{K}}+3\alpha_{\textrm{B}}}{2-\alpha_{\textrm{B}}}\right)a^{2}H^{2}+\frac{9a^{2}}{M_{*}^{2}}\left(\frac{\rho_{\textrm{m}}+p_{\textrm{m}}}{2-\alpha_{\textrm{B}}}\right)+\frac{\alpha_{\textrm{B}}k^{2}}{2-\alpha_{\textrm{B}}}\right]V_{X}\label{eq:metric_00}
\end{align}

\item Einstein (0,i)
\begin{align}
\eta^{\prime}= & \frac{3a^{2}\theta_{\textrm{m}}}{2k^{2}M_{*}^{2}}+\frac{aH}{2}\alpha_{\textrm{B}}V_{X}^{\prime}+\left[aH^{\prime}+\frac{a^{2}H^{2}}{2}\alpha_{\textrm{B}}+\frac{3a^{2}}{2M_{*}^{2}}\left(\rho_{\textrm{m}}+p_{\textrm{m}}\right)\right]V_{X}\label{eq:metric_0i}
\end{align}

\item Einstein (i,j) trace
\begin{align}
Dh^{\prime\prime}= & 2\lambda_{1}k^{2}\eta+2aH\lambda_{3}h^{\prime}-\frac{9a^{2}\alpha_{\textrm{K}}\delta p_{\textrm{m}}}{M_{*}^{2}}+3a^{2}H^{2}\lambda_{4}V_{X}^{\prime}+2a^{3}H^{3}\left[3\lambda_{6}+\frac{\lambda_{5}k^{2}}{a^{2}H^{2}}\right]V_{X}\label{eq:metric_ii}
\end{align}

\item Einstein (i,j) traceless
\begin{align}
\xi^{\prime}= & \left(1+\alpha_{\textrm{T}}\right)\eta-aH\left(2+\alpha_{\textrm{M}}\right)\xi+aH\left(\alpha_{\textrm{M}}-\alpha_{\textrm{T}}\right)V_{X}-\frac{9a^{2}\sigma_{\textrm{m}}}{2M_{*}^{2}k^{2}}\,.\label{eq:metric_ij}
\end{align}

\end{itemize}
In addition to the Einstein equations, the full system includes an
equation for the evolution of the scalar field perturbations, i.e.

\begin{align}
D\left(2-\alpha_{\textrm{B}}\right)V_{X}^{\prime\prime}+8aH\lambda_{7}V_{X}^{\prime} & +2a^{2}H^{2}\left[\frac{c_{\text{sN}}^{2}k^{2}}{a^{2}H^{2}}-4\lambda_{8}\right]V_{X}=\frac{2c_{\text{sN}}^{2}}{aH}k^{2}\eta\label{eq:metric_vx}\\
 & +\frac{3a}{2HM_{*}^{2}}\left[2\lambda_{2}\delta\rho_{\textrm{m}}-3\alpha_{\textrm{B}}\left(2-\alpha_{\textrm{B}}\right)\delta p_{\textrm{m}}\right]\,.\nonumber 
\end{align}
The definition of all these functions is

\begin{align}
D= & \alpha_{\textrm{K}}+\frac{3}{2}\alpha_{\textrm{B}}^{2}\\
\lambda_{1}= & \alpha_{\textrm{K}}\left(1+\alpha_{\textrm{T}}\right)-3\alpha_{\textrm{B}}\left(\alpha_{\textrm{M}}-\alpha_{\textrm{T}}\right)\\
\lambda_{2}= & -\frac{3\left(\rho_{\textrm{m}}+p_{\textrm{m}}\right)}{H^{2}M_{*}^{2}}-\left(2-\alpha_{\textrm{B}}\right)\frac{H^{\prime}}{aH^{2}}+\frac{\alpha_{\textrm{B}}^{\prime}}{aH}\\
\lambda_{3}= & -\frac{1}{2}\left(2+\alpha_{\textrm{M}}\right)D-\frac{3}{4}\alpha_{\textrm{B}}\lambda_{2}\\
\lambda_{4}= & \alpha_{\textrm{K}}\lambda_{2}-\frac{2\alpha_{\textrm{K}}\alpha_{\textrm{B}}^{\prime}-\alpha_{\textrm{B}}\alpha_{\textrm{K}}^{\prime}}{aH}\\
\lambda_{5}= & \frac{3}{2}\alpha_{\textrm{B}}^{2}\left(1+\alpha_{\textrm{T}}\right)+\left(D+3\alpha_{\textrm{B}}\right)\left(\alpha_{\textrm{M}}-\alpha_{\textrm{T}}\right)+\frac{3}{2}\alpha_{\textrm{B}}\lambda_{2}\\
\lambda_{6}= & \left(1-\frac{3\alpha_{\textrm{B}}H^{\prime}}{\alpha_{\textrm{K}}aH^{2}}\right)\frac{\alpha_{\textrm{K}}\lambda_{2}}{2}-\frac{DH^{\prime}}{aH^{2}}\left[2+\alpha_{\textrm{M}}+\frac{H^{\prime\prime}}{aHH^{\prime}}\right]-\frac{2\alpha_{\textrm{K}}\alpha_{\textrm{B}}^{\prime}-\alpha_{\textrm{B}}\alpha_{\textrm{K}}^{\prime}}{2aH}-\frac{3\alpha_{\textrm{K}}p_{\textrm{m}}^{\prime}}{2aH^{3}M_{*}^{2}}\\
\lambda_{7}= & \frac{D}{8}\left(2-\alpha_{\textrm{B}}\right)\left[4+\alpha_{\textrm{M}}+\frac{2H^{\prime}}{aH^{2}}+\frac{D^{\prime}}{aHD}\right]+\frac{D}{8}\lambda_{2}\\
\lambda_{8}= & -\frac{\lambda_{2}}{8}\left(D-3\lambda_{2}+\frac{3\alpha_{\textrm{B}}^{\prime}}{aH}\right)+\frac{1}{8}\left(2-\alpha_{\textrm{B}}\right)\left[\left(3\lambda_{2}-D\right)\frac{H^{\prime}}{aH^{2}}-\frac{9\alpha_{\textrm{B}}p_{\textrm{m}}^{\prime}}{2aH^{3}M_{*}^{2}}\right]\label{eq:lambda_8}\\
 & -\frac{D}{8}\left(2-\alpha_{\textrm{B}}\right)\left[4+\alpha_{\textrm{M}}+\frac{2H^{\prime}}{aH^{2}}+\frac{D^{\prime}}{aHD}\right]\nonumber \\
c_{\text{sN}}^{2}= & \lambda_{2}+\frac{1}{2}\left(2-\alpha_{\textrm{B}}\right)\left[\alpha_{\textrm{B}}\left(1+\alpha_{\textrm{T}}\right)+2\left(\alpha_{\textrm{M}}-\alpha_{\textrm{T}}\right)\right]\,.
\end{align}
Here $c_{\text{sN}}^{2}$ is the numerator of the sound speed of the
additional degree of freedom, which is
\begin{equation}
c_{\text{s}}^{2}=\frac{c_{\text{sN}}^{2}}{D}\,.
\end{equation}

\subsection{$\alpha$ Functions\label{sub:appendix_alphas}}

The time dependence of the $\alpha$ functions introduced in sec.~\ref{sub:alphas}
is uniquely specified for any particular Horndeski Lagrangian when
taken together with the background initial conditions. The relation
between $\alpha_{i}$ and the Horndeski functions $G_{i}\left(\phi,X\right)$
is given by

\begin{align}
M_{*}^{2}\equiv & 2\left(G_{4}-2XG_{4X}-\frac{H\phi^{\prime}XG_{5X}}{a}+XG_{5\phi}\right)\\
\alpha_{\textrm{M}}\equiv & \frac{d\ln M_{*}^{2}}{d\ln a}\\
H^{2}M_{*}^{2}\alpha_{\textrm{K}}\equiv & 2X\left(G_{2X}+2XG_{2XX}-2G_{3\phi}-2XG_{3\phi X}\right)\\
 & +\frac{12H\phi^{\prime}X}{a}\left(G_{3X}+XG_{3XX}-3G_{4\phi X}-2XG_{4\phi XX}\right)\nonumber \\
 & +12H^{2}X\left[G_{4X}-G_{5\phi}+X\left(8G_{4XX}-5G_{5\phi X}\right)+2X^{2}\left(2G_{4XXX}-G_{5\phi XX}\right)\right]\nonumber \\
 & +\frac{4H^{3}\phi^{\prime}X}{a}\left(3G_{5X}+7XG_{5XX}+2X^{2}G_{5XXX}\right)\nonumber \\
HM_{*}^{2}\alpha_{\textrm{B}}\equiv & \frac{2\phi^{\prime}}{a}\left(XG_{3X}-G_{4\phi}-2XG_{4\phi X}\right)+8HX\left(G_{4X}+2XG_{4XX}-G_{5\phi}-XG_{5\phi X}\right) \label{eq:aB}\\
 & +\frac{2H^{2}\phi^{\prime}X}{a}\left(3G_{5X}+2XG_{5XX}\right)\nonumber \\
M_{*}^{2}\alpha_{\textrm{T}}\equiv & 4X\left(G_{4X}-G_{5\phi}\right)-\frac{2}{a^{2}}\left(\phi^{\prime\prime}-2aH\phi^{\prime}\right)XG_{5X}\,.
\end{align}
and $\phi,X$ and $H$ are evaluated on their background solution
to give the particular time-dependence of the $\alpha$ functions
for that solution.

\bibliographystyle{utcaps}
\bibliography{hi_class}

\end{document}